\documentclass[usenatbib]{emulateapj}
\pdfoutput=1
\usepackage{graphicx}

\usepackage{tabularx}    
\usepackage{subfigure}  
\usepackage{amsmath, amssymb}
\usepackage{natbib}		
\usepackage[figuresright]{rotating}
\usepackage{hyperref}

\makeatletter

\newcommand{\Rmnum}[1]{\expandafter\@slowromancap\romannumeral #1@}
\makeatother
\newcommand\T{\rule{0pt}{2.6ex}}
\newcommand\B{\rule[-1.2ex]{0pt}{0pt}}

\begin{document}

 \title{Calibrating the Star Formation Rate at $z\sim1$ from Optical Data}

\author{Nick Mostek\altaffilmark{1}} 
\altaffiltext{1}{Space Sciences Laboratory, University of California, Berkeley, CA, 94720}
\email{njmostek@lbl.gov}
 
\author{Alison L. Coil\altaffilmark{2}}
\altaffiltext{2}{Center for Astrophysics and Space Sciences, University of California, San Diego, La Jolla, CA, 92093}
    
\author{John Moustakas\altaffilmark{2}}    
    
\author{Samir Salim\altaffilmark{3}}
\altaffiltext{3}{Department of Astronomy, Indiana University, Bloomington, IN, 47405}

\author{Benjamin J. Weiner\altaffilmark{4}}
\altaffiltext{4}{Steward Observatory, Department of Astronomy, University of Arizona, Tucson, AZ, 85721}

\begin{abstract}
We present a star-formation rate calibration based on optical data that is consistent with average observed rates in both the red and blue galaxy populations at $z\sim1$. The motivation for this study is to calculate SFRs for DEEP2 Redshift Survey galaxies in the $0.7<z<1.4$ redshift range, but our results are generally applicable to similar optically-selected galaxy samples without requiring UV or IR data. Using SFRs fit from UV/optical SEDs in the AEGIS survey, we explore the behavior of restframe $B$-band magnitude, observed [\ion{O}{2}] luminosity, and restframe (U-B) color with SED-fit SFR for both red sequence and blue cloud galaxies.  The resulting SFR calibration produces fit residual errors of 0.3 dex RMS scatter for the full color-independent sample with minimal correlated residual error in L[\ion{O}{2}] or stellar mass. We then compare the calibrated $z\sim1$ SFRs to two diagnostics that use L[\ion{O}{2}] as a tracer in local galaxies and correct for dust extinction at intermediate redshifts through either galaxy $B$-band luminosity or stellar mass. We find that a L[\ion{O}{2}] - M$_B$ SFR calibration commonly used in the literature agrees well with our calculated SFRs after correcting for the average $B$-band luminosity evolution in $L_{*}$ galaxies. However, we find better agreement with a local L[\ion{O}{2}]-based SFR calibration that includes stellar mass to correct for reddening effects, indicating that stellar mass is a better tracer of dust extinction for all galaxy types and less affected by systematic evolution than galaxy luminosity from $z=1$ to the current epoch. 
\end{abstract}
%We find that a SFR calibration can be calculated for all $z\sim1$ DEEP2 galaxies using a simultaneous fit in M$_B$ and restframe colors with residual errors that are within the SFR measurement error.
\section{Introduction}
\label{sec:Intro}

The global star formation rate (SFR) of large galaxy samples at intermediate redshift has been an area of active study for the past decade \citep{Steidel99, Hopkins06, Zhu09, Gilbank10}. Each SFR study has chosen a particular star-formation rate indicator (or combination of indicators) from a variety of possible measures. The indicators associated with star formation cover a broad range of wavelengths of the electromagnetic spectrum from x-ray to radio. The most studied measures include UV luminosity, which indicates the amount of massive star formation within the composite stellar population of a galaxy \citep{Heckman98}, and the luminosity of nebular emission lines such as H$\alpha$ and [\ion{O}{2}], which measure the amount of ionizing radiation from massive stars \citep[hearafter M06]{Kennicutt98, Kewley04, Moustakas06}. Additionally, deep observations from \emph{Spitzer} have tied the mid-IR continuum emission at $24\micron$ to compact star formation regions in local galaxies \citep{Calzetti07, Kennicutt09}.

All SFR diagnostics essentially measure the energy output of young stars either through direct observation of the UV flux or indirect observation of the ionizing luminosity through nebular emission lines or reprocessed emission from dust. Regardless of the diagnostic used, the SFR indicators have systematic effects which must be removed in order to give an accurate measure of the SFR. The primary physical effect that dominates systematic uncertainty in restframe UV/optical observations is the amount of dust extinction in the host galaxy. Galaxy dust extinction attenuates the UV continuum or emission line luminosity and biases the indicators to lower SFRs \citep{Kennicutt98, Kewley02, Kewley04}. Further, the amount of dust extinction within the star-forming galaxy population also correlates with galaxy luminosity \citep{WH96}, stellar mass \citep{Brinchmann04, Noeske07b, Elbaz07, Gilbank10}, and star-formation history \citep{Noeske07a}, presumably due to increased metallicity and dust production from massive stars and mass-loss return to their ISMs. Producing accurate dust-corrected SFRs from UV continuum or nebular emission lines requires at minimum a measure of the dust extinction through diagnostics such as the UV/optical slope \citep{Calzetti94, Kong04} or the Balmer decrement (H$\alpha$/H$\beta$) assuming case B recombination for typical galaxy environments \citep{Calzetti01}. 

Additional systematic uncertainties can also limit the SFR accuracy after dust extinction has been removed. For example, emission line luminosities from [\ion{O}{2}] and [\ion{O}{3}] depend on other nebular characteristics such as the chemical abundance and excitation states of the gas \citep{KD02}. M06 estimates that even after an ideal Balmer-decrement dust correction is applied, [\ion{O}{2}]-based SFRs have a lower error limit of 32\% for galaxies with typical metallicities and can have considerably larger error for more extreme abundances. Further, the gas can be excited from sources that are not associated with actual star formation, such as the hard-ionizing background radiation created by AGN, which can be a dominant effect in quiescent galaxies \citep{Yan06}. 

Beyond the UV and optical regime, IR measurements can be used to trace galaxy SFR \citep{Kennicutt98, Daddi07}; UV photons from massive stars are absorbed by dust within the host galaxy ISM and re-emmited at IR wavelengths. IR-based SFRs are most accurate for heavily attenuated environments such as young starbursts where dust-corrections and measurements of the UV luminosity are difficult. However, IR observations tend to underestimate the dust luminosity in normal spirals where much of the UV radiation is unabsorbed.  IR luminosities can also have systematic variations due to differing dust geometries in the host galaxy or contamination from old stellar populations, particularly in red-sequence galaxies \citep[hereafter S09]{Salim09}. Nevertheless, recent work has calibrated the SFR to IR luminosity in various bands for these dusty galaxies \citep{Zhu08, Calzetti09, Rieke09} and calibrated the $24\micron$ luminosity to H$\alpha$ luminosity to provide a measure of both the SFR and account for typical dust attenuation encountered across a broad range of blue galaxies \citep{Kennicutt09}. 

One way to mitigate these various systematic effects is to obtain integrated spectral and photometric measurements over a wide wavelength range, providing constraints on the dust extinction and independent cross-checks in SFR. In S09, galaxy template SEDs were fit to measurements spanning FUV to $K$-band wavelengths available from the All-Wavelength Extended Groth Strip International Survey (AEGIS) \citep{Davis07}. The S09 SFRs were based primarily on measurements of the UV luminosity and are combined with a Baysian analysis of stellar population synthesis models to fit the template SEDs. By simultaneously fitting across the panchromatic data and correcting for various astrophysical effects such as dust extinction, UV upturn in red galaxies, and SFR timescales, the dust-corrected SFRs are robust against large systematic errors that would typically limit SFR measurements at intermediate redshifts across galaxy types. We consider the SED SFRs fit to UV/optical measurements to be the least biased measure of SFR in red galaxies, particularly when the UV upturn is taken into account \citep{Han07}. These SED-based SFRs provide a reference for the SFR calibration presented here and will be discussed further in Section~\ref{sec:Datasets}.

The goal of this work is to take advantage of these robust SED-fit SFRs and provide a SFR calibration for optically-selected galaxy samples where wide multi-wavelength coverage or emission line data may not be available. Specifically, we wish to develop a suitable calibration using SFR data in the AEGIS field and extend that calibration to $z\sim1$ galaxies in the DEEP2 Redshift Survey (Davis, 2003) outside the AEGIS field. Because the SFR calibration is based on a large galaxy sample, the final means-tested SFR calibration will not be highly accurate for individual galaxies. However, it will reproduce the global SFR trends seen in the $z\sim1$ AEGIS sample and allow us to confidently study the galaxy environment at these redshifts with respect to SFR. In \S\ref{sec:Datasets} we describe the volume-limited galaxy sample selected from the DEEP2 survey that is matched to the S09 AEGIS SFRs. In \S\ref{sec:SFRcorr}, we investigate which combination of DEEP2 measurements, observed [\ion{O}{2}] luminosity, restframe $B$-band luminosity, or restframe color, that are most correlated with SFR and will therefore provide the best leverage in calibration. Our SFR model for these tests uses a weighted linear combination of observed parameters (not corrected for dust extinction), and we investigate which parameter combination delivers the most accurate SFR calibration for galaxies with both red and blue optical restframe colors (\S\ref{sec:MBUBcal} and \S\ref{sec:RedBlueCal}). In \S\ref{sec:OIISFRs}, we describe two empirical SFR estimators which use local [\ion{O}{2}] luminosity measurements and correct the SFRs for systematic effects via M$_B$ or stellar mass. We then compare these diagnostics to our best-fit SFR calibration at intermediate redshifts, and we present our conclusions in \S\ref{sec:Conclusions}. Throughout this work, we adjust all masses and SFRs to a \citet{Salpeter55} IMF and assume a $\Lambda$CDM cosmology ($\Omega_{M}=0.3$, $\Omega_{\Lambda}=0.7$, $h=0.7$). All magnitudes used in this study are in the AB system.

\section{Datasets}
\label{sec:Datasets}

The main sample for this work is drawn from the DEEP2 survey \citep{Davis03} where the [\ion{O}{2}] doublet emission lines have been measured with the DEIMOS spectrograph on Keck \citep{Faber03}. The DEEP2 survey contains four fields that are separated across the Northern sky for year-round observation, and one of the fields (Field 1) overlaps the AEGIS survey in the Extended Groth Strip (EGS). Within the EGS, DEEP2 targets all galaxies to R$_{AB}<24.1$ and is limited to $z<1.4$, beyond which the [\ion{O}{2}]  emission line falls outside the observed wavelength range (6500-9200\AA). In the non-EGS fields, targets selected by DEEP2 are limited by design at a lower redshift of $z>0.7$ due to an optical $BRI$ color cut. The total DEEP2 sample with good redshifts ($>95\%$ confidence, $z$ quality flag $>3$) is 31,656 galaxies. When available, the emission line equivalent widths in DEEP2 are measured using a nonlinear least-squares fit of a Gaussian to the emission line profiles. A line flux is then computed by measuring the continuum in a 20-60\AA\ window around the line and calibrating the flux to $K$-corrected $RI$ photometry. This method takes into account both the throughput and slit losses for DEEP2 galaxies of which nearly all have an effective radius less than the 1" DEIMOS slit width (Weiner et. al, 2012, in preparation; description of method in \citealp{Weiner07, Zhu09}).

To build a sample of $z\sim1$ SFRs, we match the DEEP2 galaxy sample to available panchromatic data from the AEGIS survey. The matched AEGIS sample contains a total of 5345 objects and is a subset of the total available DEEP2 spectra in the EGS field. From this sample, we obtained SFR estimates from template SEDs which are fit to UV, optical, and IR $K$-band photometry in S09. The composite templates are constructed from stellar population synthesis models of \citet{BC03} augmented with a two-component dust attenuation model of \citet{CF00} to account for dust associated with various star-formation histories and the intergalactic medium. The templates used in S09 also take into account extreme blue horizontal branch stars which are thought to be responsible for the ``UV upturn" seen in some early-type galaxies with old stellar populations \citep[see][for details of the SED models]{Salim07}. For 45\% of the S09 sample, the SED-fit SFRs have been compared with $24\micron$ measurements for both red sequence and blue cloud galaxies. The total IR luminosities are inferred from the $24\micron$ flux densities and galaxy redshifts by fitting to IR SED templates \citep{DH02}. S09 found that SED SFRs of the blue galaxy population (NUV-R$<3$) were consistent with the measured IR luminosities to 0.3 dex RMS. The SFRs for red galaxies (NUV-R$>5$) were less well matched in IR luminosity due to the increasing contribution of light from old stellar populations in the IR measurements. Because S09 accounts for the latter systematic effect though template fitting,  the S09 SFRs based on UV SEDs are a better estimate of the true SFR than the $24\micron$ measurements in red galaxies. 

The stellar masses for our galaxy sample are computed from the \citet{Bell03} color-M/L relations, modified to DEEP2 redshifts through the \citet{Weiner09} calibration.  The color-M/L calibration is a linear set of equations that require several measured quantities, include M$_{B}$,  restframe (U-B) and (B-V) colors, and galaxy redshift.  Independently, we confirm that our calculated stellar masses from these relations agree with stellar masses derived from $K$-band measurements \citep{Bundy06} to 0.3 dex RMS with no systematic offset. As an additional consistency check, we find that the color-M/L stellar masses agree with those generated by the more sophisticated S09 SED-fitting techniques to within 0.3 dex. We use stellar masses from the M/L-color calibration rather than stellar masses obtained from the SED fitting because we want the input parameters in our SFR calibration to be easily reproducible and independent from the SED fitting that yields our fiducial SFRs. 

\subsection{Sample selection}
To produce a statistical sample volume-limited in M$_{B}$ for both red and blue galaxies, we restrict the matched DEEP2 / AEGIS galaxy sample to M$_{B}<-20$ for $0.74<z<1.0$, resulting in 1039 total galaxies. The selected redshift range and galaxy luminosity limit is similar to color-independent, volume-limited DEEP2 samples used in previous environmental studies \citep{Cooper06, Gerke07, Coil08}. Approximately 17\% of the selected red galaxies and 9\% of the blue galaxies in our sample have no measured amount of L[\ion{O}{2}].  The L[\ion{O}{2}] measurement failures in blue galaxies are likely due to interactions of the emission line with bright sky emission lines, which happens at an nearly equal rate for both blue and red galaxies over this redshift range. The incompleteness due to sky line interaction is generally independent of the galaxy restframe $B$-band luminosity, color, and redshift. Removing the galaxies without measured [\ion{O}{2}] reduces the sample size to 885 DEEP2 galaxies matched with AEGIS and restricts our SFR calibration analysis to galaxies with a SED-fit SFR limit of log($\psi$)$>-1.0$  and [\ion{O}{2}] line luminosity limit of log(L[\ion{O}{2}])$>39.7$ at $z\sim1$.\footnote{The SFR $\psi$ is in units of M$_{\sun}$ yr$^{-1}$, and L[\ion{O}{2}] is in ergs s$^{-1}$ throughout this study.}

While blue galaxies in our sample generally have well detected L[\ion{O}{2}] measurements ($>3\sigma$), a significant fraction of the red galaxies with L[\ion{O}{2}] measurements have large error due to inherently lower line luminosities. Restricting the L[\ion{O}{2}] measurements to better than 3$\sigma$ in DEEP2
reduces the sample to 80\% of the volume-limited blue galaxies and 27\% of red galaxies. For comparison, \citet{Yan06} found that 35\% of all volume-limited red galaxies in SDSS had [\ion{O}{2}] detected at a $3\sigma$ level in EW. Because we are interested in testing SFR calibrations for a uniform sample with both L[\ion{O}{2}]  and M$_B$, we wish not to limit the completeness in M$_B$, which is well measured for all galaxies, due to the higher L[\ion{O}{2}] errors.  Therefore, we opt to keep all galaxies with measured L[\ion{O}{2}] in our sample and weight the SFR fits by the measurement errors.  

\begin{figure}[tb]
\centering
\includegraphics[width=0.98\columnwidth]{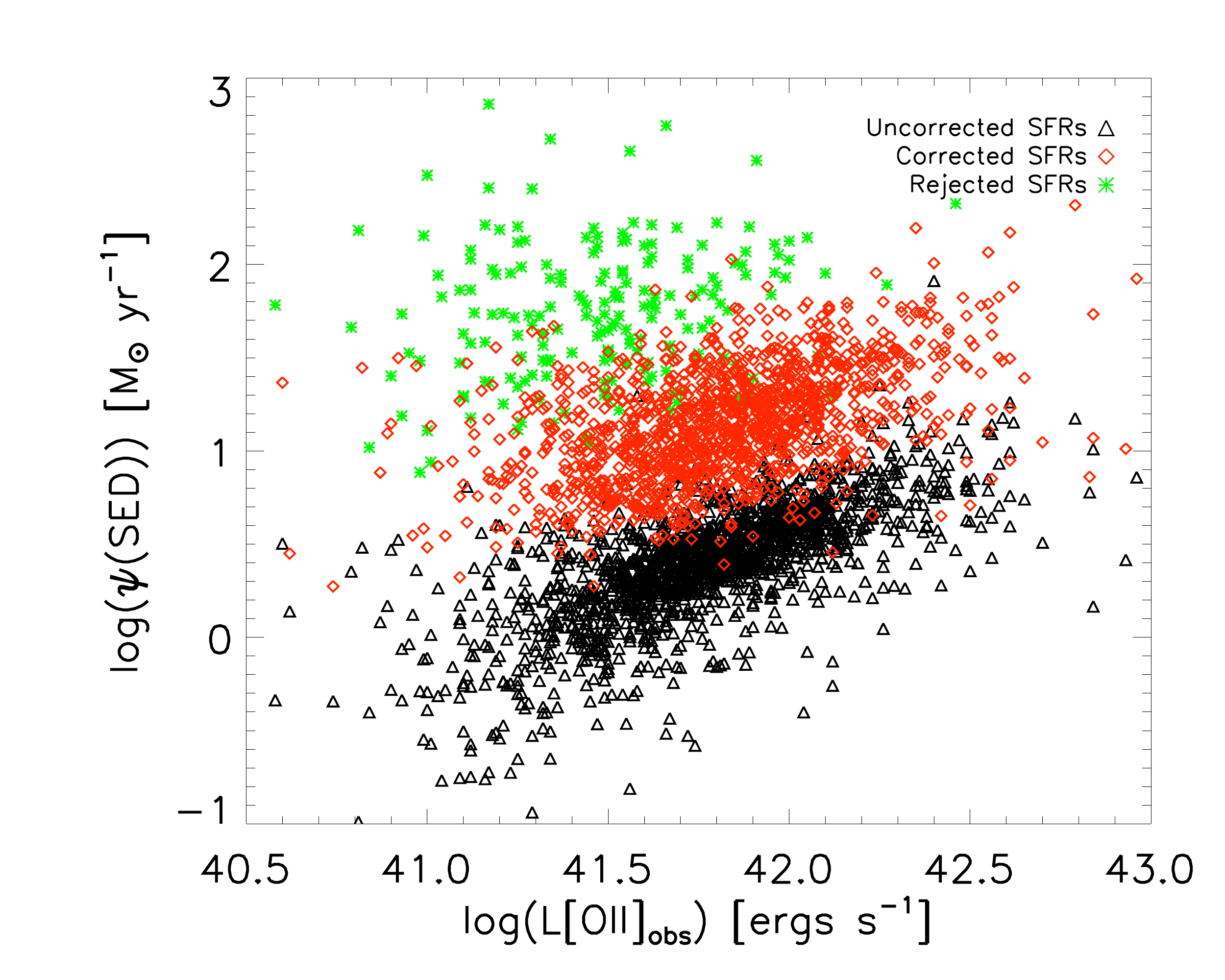}
\caption{Comparison of the S09 SFRs for blue galaxies from $0.74<z<1.4$ with observed L[\ion{O}{2}] in DEEP2. The plotted points are for both uncorrected (black triangles) and dust corrected (red diamonds) SFR values. The green stars denote the rejected S09 SFRs for heavily dust-extincted blue galaxies that require a factor $>20$ in dust correction. Note that the uncorrected SFRs are tightly correlated with observed L[\ion{O}{2}], indicating that the dust correction introduces systematic uncertainty into the corrected SFR values.}
\label{fig:SFRreject}
\end{figure}

We place one final restriction on the sample to remove galaxies that have a large dust-corrected SFRs in the S09 data. Figure~\ref{fig:SFRreject} shows the dust-corrected and uncorrected S09 SFRs in the matched DEEP2 / AEGIS sample as a function of observed [\ion{O}{2}] luminosity from DEEP2.  We find that the uncorrected UV/optical SED SFRs and observed L[\ion{O}{2}] are tightly correlated due to similar dust extinction properties in the UV. After the dust correction is applied, a small population (13\%) of galaxies have SFRs that are $>3\sigma$ outliers from the mean dust-corrected distribution. These ``dusty" star-forming galaxies have SFR corrections of more than a factor of 20 and are noted in S09 as primarily intermediate color (green valley) galaxies. Since we are attempting to calibrate SFRs at $z\sim1$ with limited optical observations, we assume that the dust extinction for these obscured galaxies cannot be accurately measured on an individual galaxy basis (usually done through IR measurements) and remove them from the following analysis. While removing the outliers allows our calibrations to more closely track the ``typical" galaxy in our sample, this underlying assumption restricts the SFR calibrations in this study to galaxy samples with moderate dust extinction and we acknowledge that our results will not be accurate for galaxies that require extraordinarily large dust corrections. The sample used in our SFR calibration is reduced to a total of 771 galaxies after we remove these galaxies.

\begin{figure}[tb]
\centering
\includegraphics[width=0.98\columnwidth]{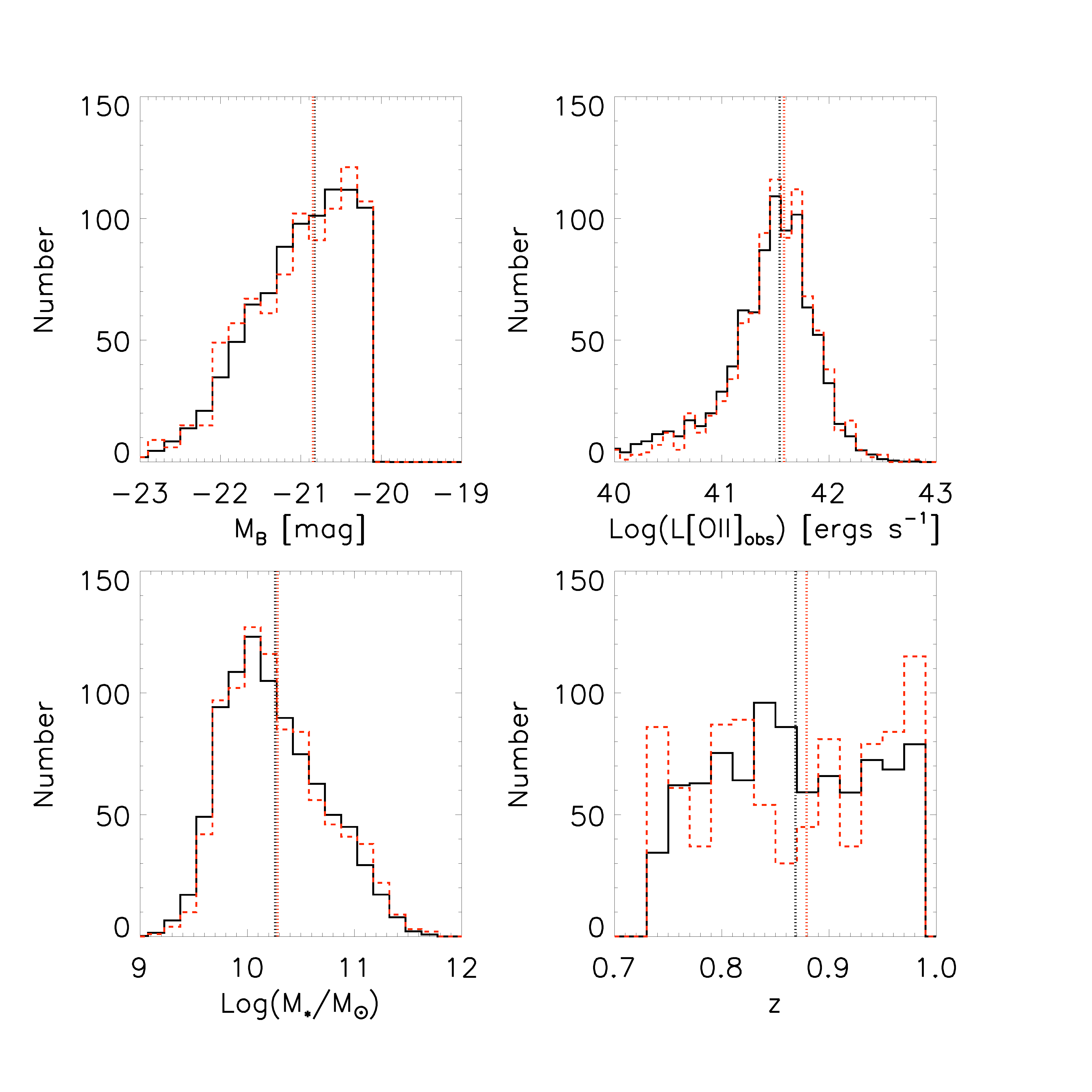}
\caption{Histograms of the selected and matched AEGIS / DEEP2 sample from $0.74<z<1.0$ (red, dashed) and the entire selected DEEP2 galaxy sample over the same redshift range in the EGS (black, solid).  The DEEP2 EGS sample histograms have been normalized to the total number in the AEGIS matched sample to show the relative shape of the distributions. The dotted vertical lines mark the median values for each sample. }
\label{fig:SampleHists}
\end{figure}

To confirm that our selected and matched AEGIS / DEEP2 sample is a representative subset of optically selected $z\sim1$ galaxies, we compare the measured parameter distributions for the matched AEGIS / DEEP2 sample to the full EGS DEEP2 galaxy sample using the same selection cuts. This test must be restricted to the EGS field because the original DEEP2 target selection within the EGS was color-independent, whereas the three other DEEP2 fields were color selected.  Figure~\ref{fig:SampleHists} shows the selected sample histograms for [\ion{O}{2}] line luminosity, restframe B-band luminosity, stellar mass, and redshift. Because the matched and selected galaxy sample is roughly 10\% of the entire DEEP2 EGS sample, the DEEP2 histograms have been normalized to the total matched AEGIS sample for easy comparison of the distributions. We find that the AEGIS / DEEP2 matched sample is an accurate subset of volume-limited, optically-selected galaxies in this redshift range and produces equivalent median values as the full EGS DEEP2 sample with log(L[\ion{O}{2}])=41.5, M$_{B}=-20.8$, log(M$_{*}$/M$_{\sun}$)=10.3 and redshift $z=0.88$.

\section{SFR correlations}
\label{sec:SFRcorr}

\subsection{Trends in blue and red galaxies}
\label{sec:Galtrends}
\begin{figure*}[tb]
\centering
\includegraphics[height=3.0in]{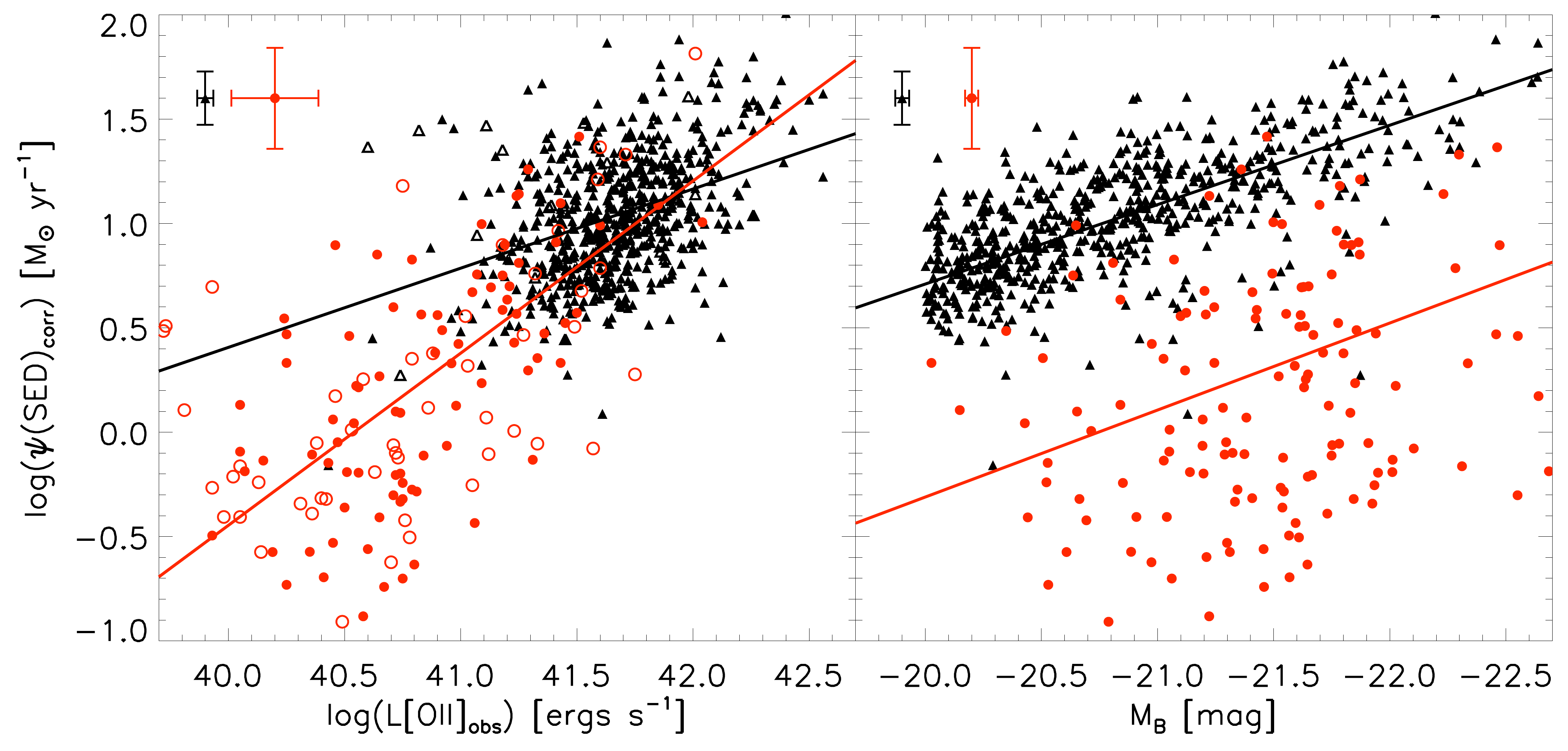}
\caption{Correlated trends in SFR with respect to observed L[\ion{O}{2}] and M$_B$ for the restframe red (red circles) and blue (black triangle) galaxies in our matched sample. The SFRs used in this study are computed from UV-optical data fit to SED templates in S09 and the linear trends are weighted by the measurement errors. The points with error bars in the upper left of each plot show the median error for the color-separated galaxy measurements. In the case of L[\ion{O}{2}], open symbols mark the galaxies where the [\ion{O}{2}] luminosity has less than a $2\sigma$ measurement in DEEP2. }
\label{fig:SFRtrends}
\end{figure*}

In this study, we explore the systematic correlation between the dust-corrected SFR and three measured physical parameters: 1) observed [\ion{O}{2}] line luminosity, 2) restframe $B$ magnitude, and 3) restframe (U-B) color. These parameters are based $only$ on optical measurements (i.e. no UV or IR data iss required) and therefore may be readily produced over wide fields where spectroscopic or photometric redshifts exist. The correlated trends of observed L[\ion{O}{2}] and M$_B$ with SFR are shown in Figure~\ref{fig:SFRtrends}, where we have separated the least-squares fits of both red and blue galaxies using the restframe color criteria of \citet{Willmer06}. The linear fits are weighted by the measurement error in both the independent and dependent variables plus an additional ``intrinsic scatter" term added in quadrature (see Section~\ref{sec:FitMethod} for more detail). In the case of L[\ion{O}{2}], we plot the measurements with $<2\sigma$ significance using an open circle to indicate that they are less important in the fit.

Several trends are apparent in the figure: \emph{i}) (U-B) color clearly separates the red galaxies with lower SFRs from the blue galaxies with higher SFRs, \emph{ii}) the red galaxies clearly have a greater scatter about the mean regression line in both observed L[\ion{O}{2}] and M$_B$ relative to the blue galaxies, and \emph{iii}) the red and blue galaxies have similar SFR slopes in M$_B$ but have different slopes in L[\ion{O}{2}]. The similar SFR slope between red and blue galaxies and restframe M$_B$ might indicate that the galaxy $B$-band luminosity is a more accurate indicator of the corrected SFR regardless of galaxy color (albeit with a small offset between SFR zeropoints), but the SFR variation among the red galaxy population is too large for a definitive statement. Figure~\ref{fig:SFRtrends} also shows that there is a large variation in SFR for blue galaxies of a fixed L[\ion{O}{2}], demonstrating that observed [\ion{O}{2}] line luminosities with small measurement errors may still have large systematic variations in corrected SFR due to the extinction effects. 

The difference in the SFR scatter of red and blue galaxies relative to their mean trends in M$_B$ and L[\ion{O}{2}] occurs for several reasons. Red galaxies are typically quiescent in star formation and have inherently lower UV flux than that of the blue galaxies, leading to higher measurement uncertainty in red galaxy SED SFRs. Similarly, the measurements in L[\ion{O}{2}] have higher measurement error in red galaxies simply because the emission line fluxes are fainter than blue galaxies. For red galaxies, we also expect the measurements are biased by [\ion{O}{2}] emission not associated with star formation. \citet{Yan06} showed that L[\ion{O}{2}] is dominated by AGN emission for a significant fraction of the local red galaxy population and that this emission is not representative of true star formation. We will study the extent to which the observed L[\ion{O}{2}] can be used as a direct proxy for quiescent galaxy SFR in the following section.

\subsection{Fitting Methodology}
\label{sec:FitMethod}
To obtain an empirical SFR calibration for our matched DEEP2 / AEGIS sample, we perform a weighted $\chi^2$ fit using observed L[\ion{O}{2}] and M$_B$, where the galaxies have been separated into red and blue subsamples by restframe (U-B) color. The galaxy SFR fits follow a simple linear combination model according to 
\begin{equation}
\psi_{model}= C_0 + \sum_{i=1}^n C_iP_i,
\label{eq:fitmodel}
\end{equation}
where $n$ is the number of observed parameters for the linear model and $C_i$ is the scalar fit coefficient for the measured parameter $P_i$. The linear SFR model is then fit to the corrected S09 SFRs, $\psi_{S09}$, by minimizing
\begin{equation}
\chi^2=\sum_{k=1}^m \frac{(\psi_{model} - \psi_{S09})_k^2}{(\sigma_{o}^2+\sigma_{i}^2)_k},
\label{eq:chisquare}
\end{equation}
where $k$ is the index of $m$ sample galaxies. The $\chi^2$ is weighted by the quadrature sum of the observed parameter variance, $\sigma_{o}^2$, and the intrinsic scatter in the distribution, $\sigma_{i}^2$. Typically, the error in $\psi_{S09}$ dominates the measured variance when compared to the observed independent variables of M$_B$ and L[\ion{O}{2}]. However, in the case of the low [\ion{O}{2}] luminosity in red galaxies, it is important to include the luminosity error to properly weight the fit. The intrinsic scatter is computed through a bootstrap method by varying the $\sigma_{i}$ term until the reduced $\chi^2/NDF=1$, ensuring that the computed fit coefficient errors are properly scaled and that the fit is unbiased given differing errors in the dependent and independent variables. We refer the reader to the appendix of \citet{Weiner06} for details of weighting fits with intrinsic scatter.

Once the $\chi^2$ fit is minimized, we subtract the best-fit SFR from the matched S09 SFR and measure the RMS of the residual errors. We also judge the performance of the calibration by measuring the correlation between residual SFR errors and the measured parameters of M$_B$, L[\ion{O}{2}], and stellar mass. A large residual correlation is an indication that the calibration process has not fully characterized the SFR distribution and that a better calibration may be sought. Because stellar mass and restframe colors are correlated by color-M/L relation, we expect that any residual systematic correlated error in color from our SFR calibration will also produce a residual error correlated with the calculated stellar masses.

\subsection{Color-separated Empirical SFR Calibration}
\label{sec:RedBlueCal}
\begin{figure*}[tb]
\centering
\includegraphics[width=6.5in]{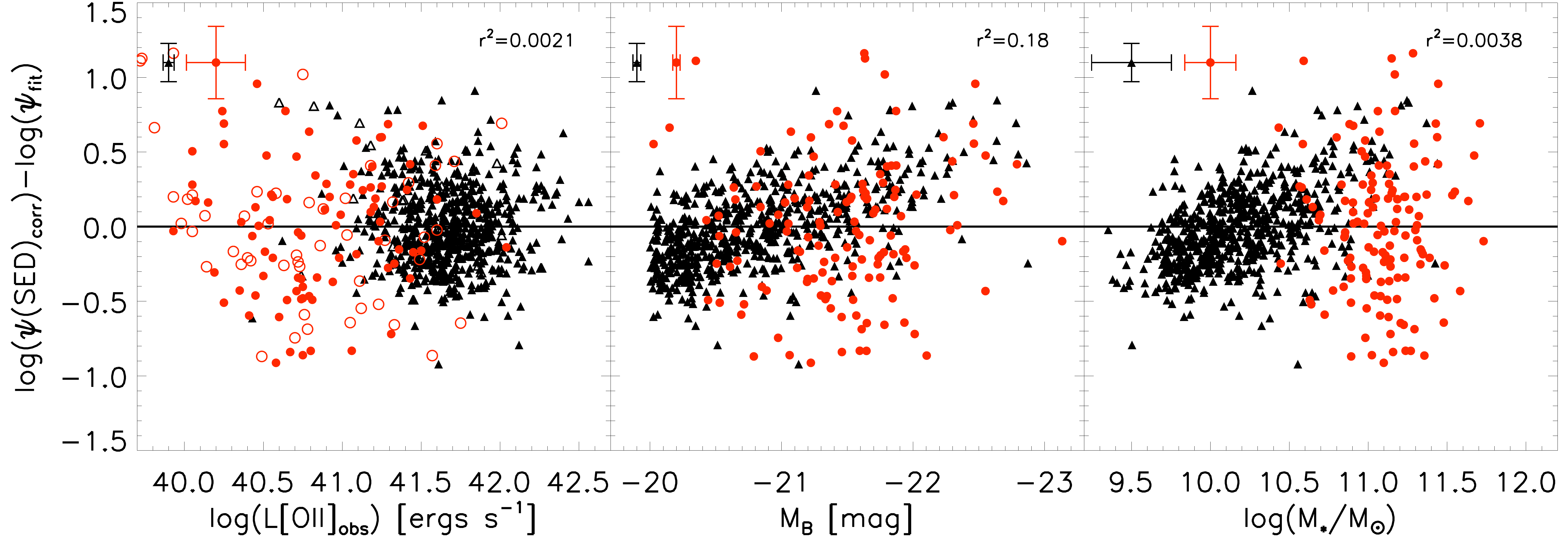}
\caption{SFR fit residuals using L[\ion{O}{2}] as the independent parameter in the SFR calibration. Fits for red (circles) and blue (triangles) galaxies are derived separately. The points with error bars in the upper left of each plot shows the median errors for each galaxy color, and the Pearson correlation coefficient values for the combined sample are shown in the upper right. In the case of stellar mass, the error on log(M$_*$/M$_{\sun}$) is calculated from the RMS of the DEEP2 masses to the fit S09 masses and therefore accounts for the systematic error in estimating stellar mass from the color-M/L relation.}
\label{fig:OIIbluered}
\end{figure*}

As discussed in Section~\ref{sec:Galtrends} and shown in Figure~\ref{fig:SFRtrends}, we initially separate the matched sample by restframe blue and red color and fit the linear model of Equation~\ref{eq:fitmodel} with  L[\ion{O}{2}] and M$_B$ as the independent variables. We report the best-fit coefficients and parameter errors of this calibration in Table~\ref{tab:SFRresults}. In Figure~\ref{fig:OIIbluered}, we show that by fitting only L[\ion{O}{2}] as an independent parameter in each galaxy type, we can achieve a SFR fit that produces residual errors with an RMS of 0.27 dex scatter for blue galaxies, 0.46 dex for red galaxies, and 0.31 dex for all galaxies in the sample. This result suggests that if restframe colors and [\ion{O}{2}] measurements are available, L[\ion{O}{2}] can be a decent tracer of the average SFR in large galaxy samples even though AGN activity may be contributing to the [\ion{O}{2}] flux in individual red galaxies.  However, there remains a residual systematic correlation along M$_B$ after calibration, as indicated by the Pearson correlation coefficient of $r^2=0.18$. The residual trend indicates that the full behavior of SFR has not been calibrated with L[\ion{O}{2}] alone and that some improvement can still be made to reduce the scatter in the calibrated SFRs. This residual correlation is likely due to the degenerate effects of dust reddening on the observed [\ion{O}{2}] luminosity, $B$-band galaxy luminosity, and restframe color.

\begin{figure*}[tb]
\centering
\includegraphics[width=6.5in]{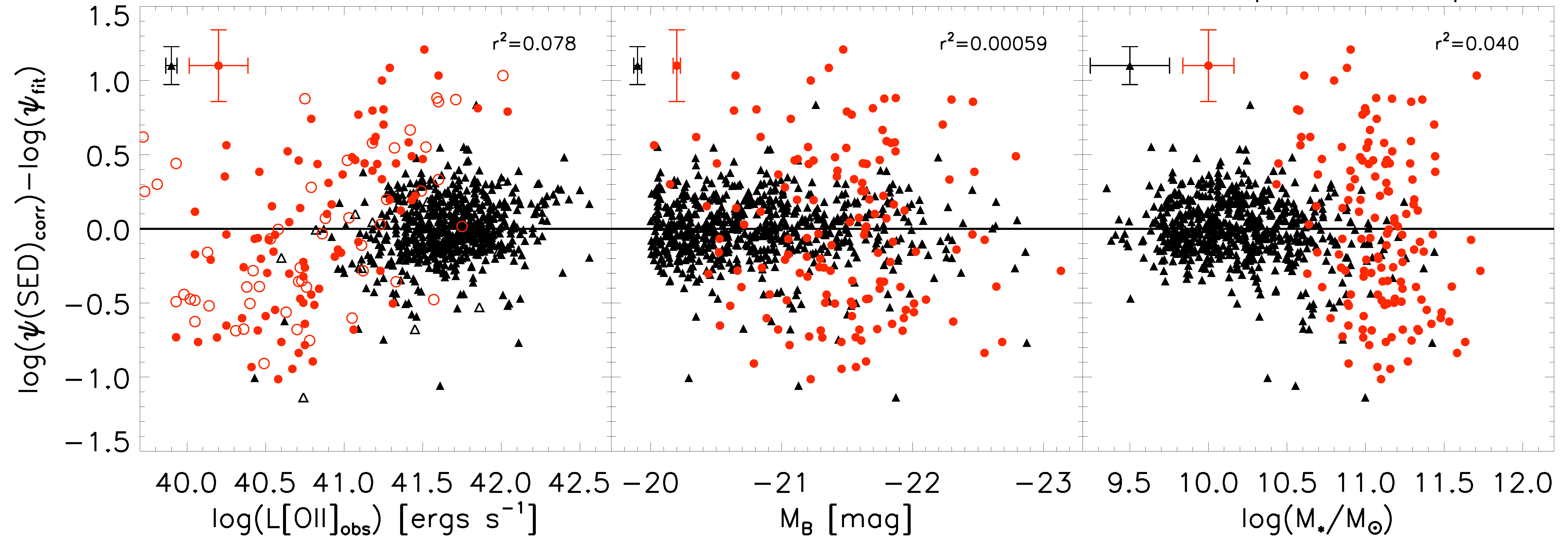}\\
\vspace{0.1in}
\includegraphics[width=6.5in]{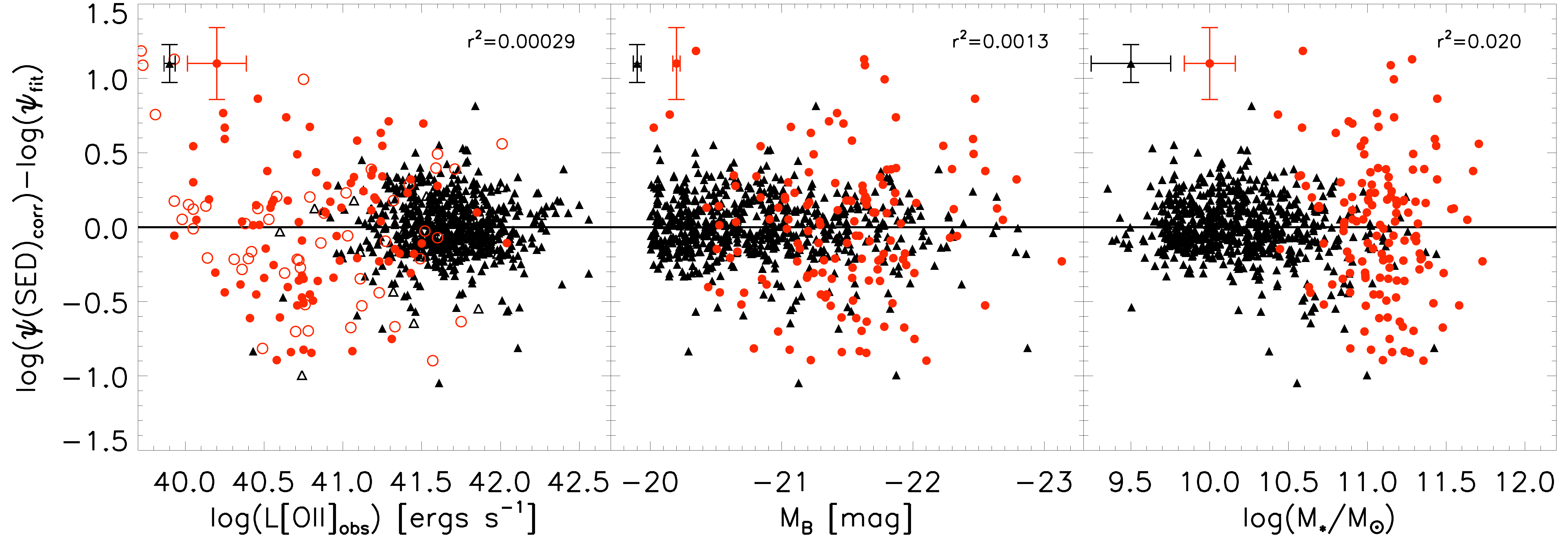}\\
\caption{SFR fit residuals using only M$_B$ (top) and both L[\ion{O}{2}] and M$_B$ (bottom) as the independent parameters in the SFR calibration. Fits for red and blue galaxies are derived separately.}
\label{fig:MBOIIbluered}
\end{figure*}

Similarly, we obtain a SFR calibration by fitting the linear trends in M$_B$ separately for red and blue galaxies (top plot of Fig~\ref{fig:MBOIIbluered}) and the fit coefficients are tabulated in Table~\ref{tab:SFRresults}. The blue galaxy SFRs are highly correlated with $B$-band luminosity (Fig.~\ref{fig:SFRtrends}), which provides greater leverage in the calibration over L[\ion{O}{2}]. The fit SFR residual errors from the M$_B$ fit show a reduced scatter for blue galaxies (0.21 dex), an increased scatter for red galaxies (0.54 dex), and no change in the scatter for the combined sample (0.30 dex). There is less correlation in M$_B$ (by construction) and stellar mass, but we observe that the SFR residual errors are now correlated with L[\ion{O}{2}], particularly for the red galaxies.  Again, the residual correlations in L[\ion{O}{2}] after calibrating with M$_B$ likely results from the differing amount of dust extinction inherent in these two observed, uncorrected measures.

\begin{table*}[tb]
\caption{Summary of best fit coefficients for SFR calibration separated by galaxy color}
\begin{center}
\begin{tabularx}{0.89\textwidth}{cccccccc}
\hline
\hline
\multicolumn{2}{c}{Fit Parameters\T} & \multicolumn{3}{c}{Blue Galaxies} & \multicolumn{3}{c}{Red Galaxies} \\
$P{_1}$ & $P_{2}$ & $C_{0,b}$ & $C_{1,b}$  &  $C_{2,b}$ &  $C_{0,r}$ & $C_{1,r}$  & $C_{2,r}$  \\
\hline
L[\ion{O}{2}]$^a$\T & --~~&~~0.722 $\pm$ 0.028 & 0.470 $\pm$ 0.038 & --~~&~~0.351 $\pm$ 0.040 & 0.763 $\pm$ 0.077 & --\\[3pt]
M$_{B}$$^b$ & --~~&~~1.099 $\pm$ 0.008 & -0.356 $\pm$ 0.013 & --~~&~~0.065 $\pm$ 0.064 & -0.304 $\pm$ 0.083 & --\\[3pt]
L[\ion{O}{2}]~~\B& M$_B$~~&~~1.003 $\pm$ 0.024 & 0.143 $\pm$ 0.031 & -0.346 $\pm$ 0.014~~&~~0.302 $\pm$ 0.055 & 0.738 $\pm$ 0.081 & -0.088 $\pm$ 0.069 \\[3pt]
\hline
\hline
\end{tabularx}
\end{center}
\hspace{0.8in}$^a$ The [\ion{O}{2}] luminosity has a zeropoint of L[\ion{O}{2}]=$10^{41}$ ergs s$^{-1}$ cm$^{-2}$. 
\vspace{-0.00in}

\hspace{0.8in}$^b$ The $B$-band luminosity has a zeropoint of M$_B$=-21. \\
\label{tab:SFRresults}
\end{table*}
\begin{table}[thb]
\caption{Intrinsic scatter and residual errors for SFR calibrations separated by galaxy color$^a$}
\begin{center}
\begin{tabularx}{0.98\columnwidth}{cccccccc}
\hline
\hline
\multicolumn{2}{c}{Parameters\T}  & \multicolumn{2}{c}{Intrinsic Scatter} & $\Delta_{mean}$ & \multicolumn{3}{c}{Residual RMS} \\
$P{_1}$ \B& $P_{2}$ &~~blue & red & all & blue & red & all \\
\hline
L[\ion{O}{2}]\T & -- &~~0.22 & 0.29 & 0.00 & 0.27 & 0.46 & 0.31 \\[3pt]
M$_{B}$ & -- &~~0.15 & 0.49 & -0.01 & 0.21 & 0.54 & 0.30 \\[3pt]
L[\ion{O}{2}] & M$_B$ &~~0.13 &  0.29 & -0.01 & 0.21 & 0.45 & 0.27 \\[3pt]
\hline
\hline
\end{tabularx}
\end{center}
\vspace{0.04in}$^a$ SFR errors are in units of dex.
\label{tab:SFRresids}
\end{table}
We can also combine both M$_B$ and L[\ion{O}{2}] measurements to reduce the residual SFR correlations below what can be obtained by fitting each independent parameter alone. The combined calibration fit achieves an RMS scatter of 0.21, 0.45, and 0.27 dex for the blue, red, and combined galaxy sample, respectively (c.f. bottom of Fig.~\ref{fig:MBOIIbluered}).  The inclusion of L[\ion{O}{2}] with M$_B$ actually provides constraint on the \emph{red} galaxies, reducing the scatter from 0.54 to 0.45 dex. This result may seem counterintuitive since the measurement error of L[\ion{O}{2}] is larger for red galaxies than blue galaxies and AGN are known to contribute significantly the [\ion{O}{2}] luminosity.  However, it is worth noting that galaxies with red restframe colors will span larger range of dust extinction values and SFR, from elliptical galaxies with low dust content and low dust-corrected SFR values to late-type galaxies with heavy dust extinction and high dust-corrected SFR. Even with significant measurement error, we find that the additional information from L[\ion{O}{2}] to M$_B$ helps provide a valuable constraint on the dust-corrected SFR in red galaxies.  The best fit coefficients for all three color-separated SFR calibrations presented in this section are summarized in Table~\ref{tab:SFRresults} and the residual RMS scatter and mean residual offsets are reported in Table~\ref{tab:SFRresids}.

\subsection{Constructing a single SFR calibration}
\label{sec:MBUBcal}
While separating the red and blue galaxy populations enables a SFR calibration across galaxy types with minimal residual correlation and scatter, it inherently requires all three measurement parameters considered in this study (L[\ion{O}{2}] , M$_B$, and restframe color) to utilize the relation. Ideally, one would like a single, simple relation to calibrate all galaxies to the SED SFRs while minimizing the number of required measurement parameters.  As shown in Section~\ref{sec:Galtrends}, the red and blue galaxy SFRs clearly have differing trends in the observed parameters and the intrinsic color dependence cannot be ignored. For example, a SFR calibration using L[\ion{O}{2}] for all galaxies \emph{without} color information produces a 35\% increase in the SFR fit residual scatter with large residual correlations in L[\ion{O}{2}], M$_B$, and M$_{*}$. As an alternative to separating galaxies by restframe color, we can add restframe color information directly into the SFR calibration and fit the entire galaxy sample simultaneously. Including the restframe color as an independent parameter in the fit produces a continuous SFR calibration between galaxy types and helps break the degeneracies between the galaxy color, SFR, and dust reddening. 

\begin{figure*}[thb]
\centering
\includegraphics[width=6.5in]{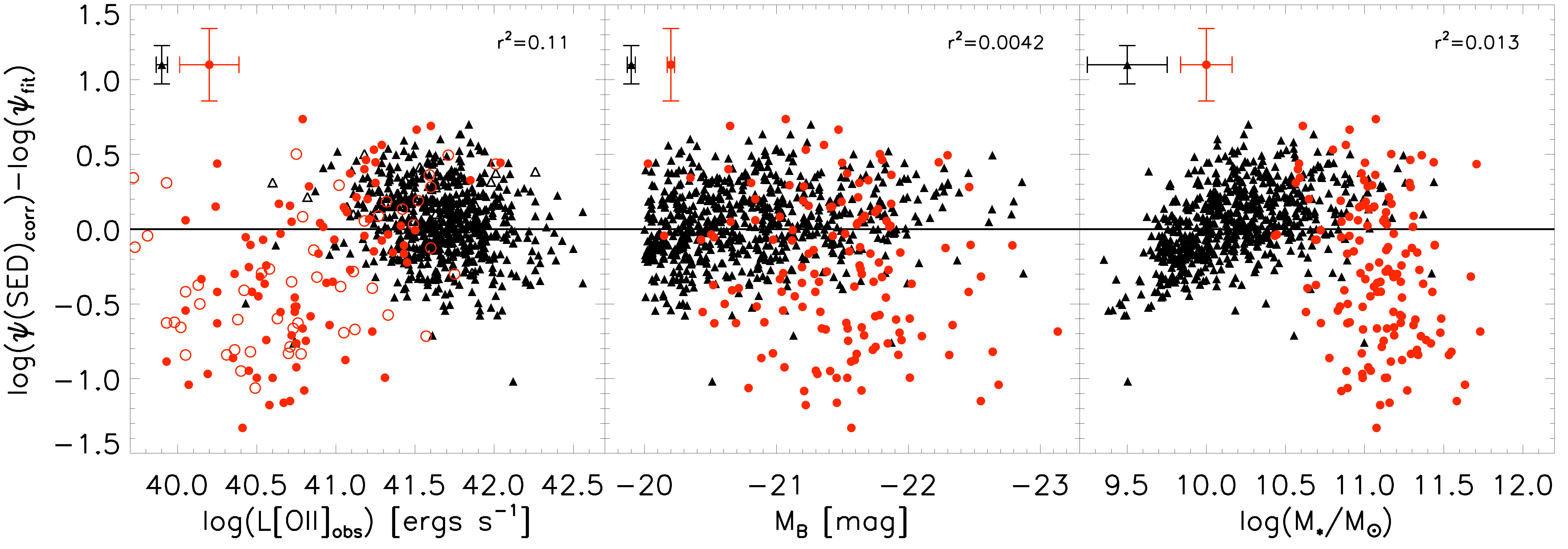}\\
\vspace{0.1in}
\includegraphics[width=6.5in]{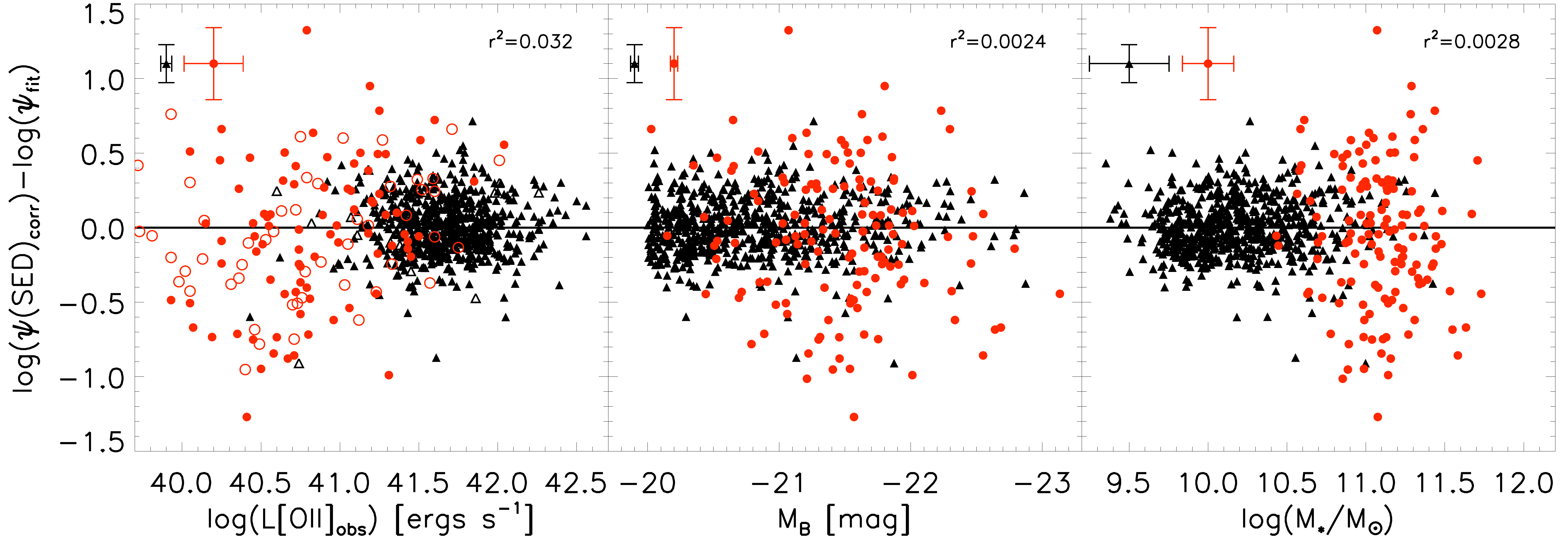}\\
\caption{SFR fit residuals using M$_{B}$ and (U-B) as the dependent parameter in the SFR calibration for all galaxies. The top figure shows the fit residuals from a linear fit in both independent variables, while the bottom figure shows the result of a second-order polynomial in (U-B).}
\label{fig:MBUB}
\end{figure*}

Figure~\ref{fig:SFRtrends} demonstrated that the dust-corrected SED SFRs are highly correlated with galaxy $B$-band luminosity, which provides excellent calibration leverage for blue galaxy SFRs and marginal leverage for red galaxy SFRs. We also observed that the correlation between SFR and M$_B$ between red and blue galaxies may have similar slopes with only a difference in the zeropoint of the relation. Motivated by these observed trends, we use the form of Eq.~\ref{eq:fitmodel} and perform a fit with both restframe M$_B$ and (U-B) color for all galaxies; the resulting residual errors are shown in Figure~\ref{fig:MBUB} (top plot).  We find that the SFR behavior of the full galaxy sample is not well described by a simple linear fit in M$_B$ and (U-B) alone.  The SFR calibration produces residual errors with $\sigma_{all}=0.33$ dex scatter for the full sample and large residual error correlated in M$_B$ and stellar mass. The residual correlations are apparently non-linear, particularly in the case of stellar mass, and therefore are not fully characterized by the Pearson correlation coefficient alone.  

Because the fit residuals in the M$_{B}$ and (U-B) color calibration are seemingly non-linear, we modify our linear SFR calibration model (Eq.~\ref{eq:fitmodel}) to include a second-order term in (U-B) (bottom plot of Figure~\ref{fig:MBUB}).  The inclusion of this non-linear color term produces an better fit to the dust-corrected SED SFRs for the full galaxy sample. The final RMS scatter in the SFR fit residuals are  $\sigma_{blue}=0.19$ dex, $\sigma_{red}=0.47$ dex, and $\sigma_{all}=0.26$ dex. The fit residuals from this calibration are \emph{statistically} better than all other methods considered in this study,  and the calibration does not require L[\ion{O}{2}] to be included in the fit as was necessary in \S\ref{sec:RedBlueCal}. Further, the SFR calibration leaves no significant correlations in M$_B$, M$_{*}$, \emph{and} L[\ion{O}{2}] (r$^{2}<0.032$). 

\begin{figure}[tb]
\centering
\subfigure{\includegraphics[width=0.98\columnwidth]{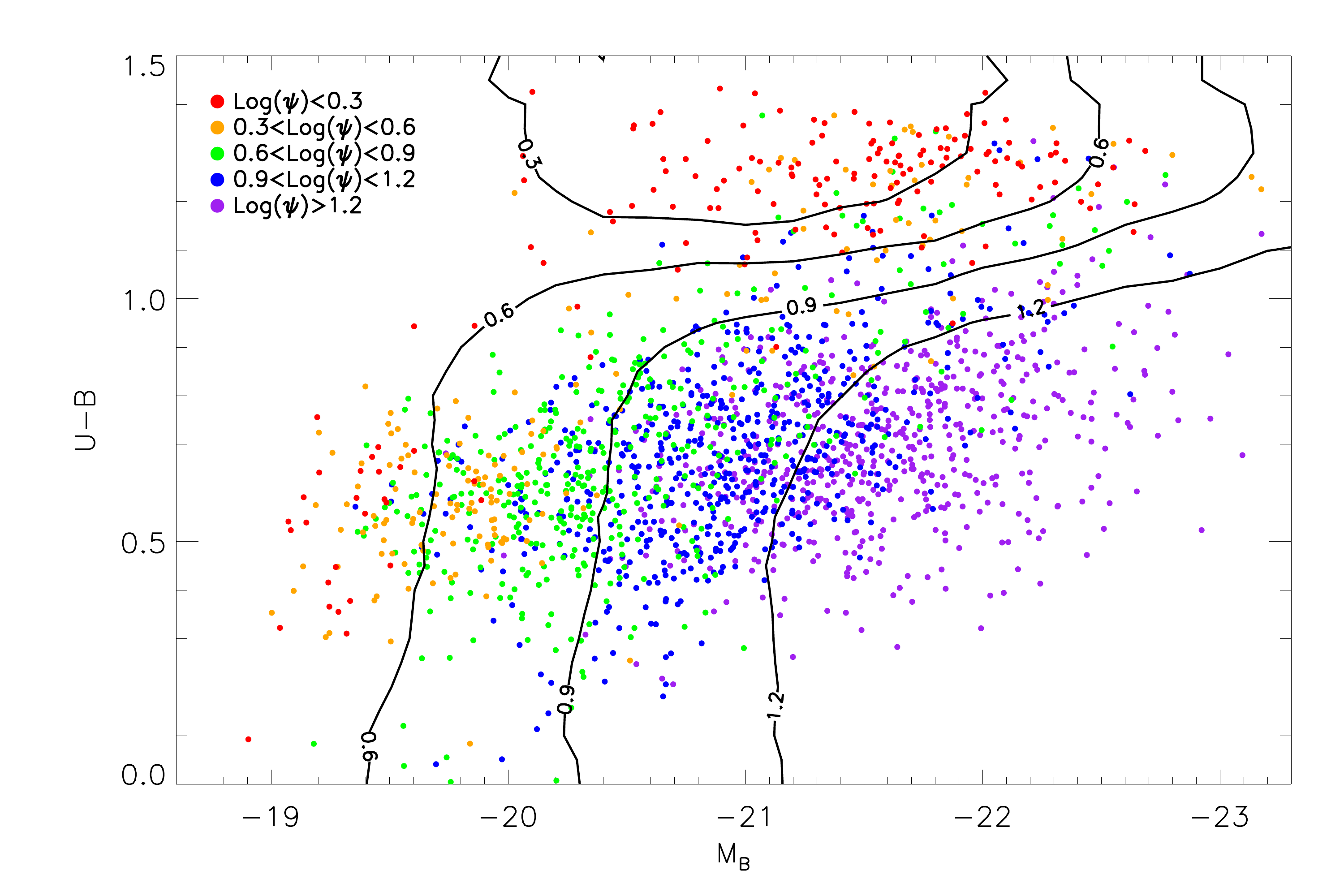}}
\caption{The M$_{B}$ - (U-B) plane with SFRs measured in the AEGIS survey (S09) and matched to the DEEP2 catalog from $0.74<z<1.4$.  The contour lines correspond to average SFRs of log($\psi$)=[0.3, 0.6, 0.9, 1.2] M$_{\sun}$ yr$^{-1}$.}
\label{fig:S09SFRcontours}
\end{figure}

The reason for the poor fit using only linear terms in M$_B$ and (U-B) alone can be seen by studying the behavior of the galaxy SFRs in the M$_B$ - (U-B) color plane.  Figure~\ref{fig:S09SFRcontours} shows the S09 SFRs matched to sources within our DEEP2 sample from $0.74<z<1.4$; the overplotted contours correspond to lines of average SFR with log($\psi$)=[0.3, 0.6, 0.9, 1.2] M$_{\sun}$ yr$^{-1}$.  The SED-fit SFR behavior are decidedly non-linear in M$_B$ and (U-B) color. For the blue galaxies, the SFRs primarily increase with M$_{B}$ with only a weak dependence on (U-B) color. This SFR behavior in the blue cloud is in agreement with measured SFRs at $z=1.4$ \citep{Weiner09} and is consistent with the assumption that the UV slope and (U-B) colors are correlated, therefore associating M$_B$ with UV flux. However, as star formation trends from active to quiescent around (U-B)=1, the SFR contours rapidly transition to much lower SFRs and have a weaker dependence on M$_{B}$. 

In addition to the differing SFR trends in M$_{B}$ between red and blue galaxies, the differential extinction between independent fit parameters, such as the uncorrected $B$-band luminosity and observed [\ion{O}{2}] luminosity, can couple the fit parameters such that independent linear coefficients are not an accurate description of the global SFR trends.  Using an additional second-order (U-B) color term in the SFR calibration allows an extra degree of freedom within the fit to simultaneously correct for both intrinsic galaxy color between red and blue galaxies and dust reddening effects.  By separating the degenerate color effects with an additional color fit parameter, we can effectively reduce the systematic coupling in the SFR fit residuals. 

\begin{figure*}[tb]
\centering
\includegraphics[width=6.5in]{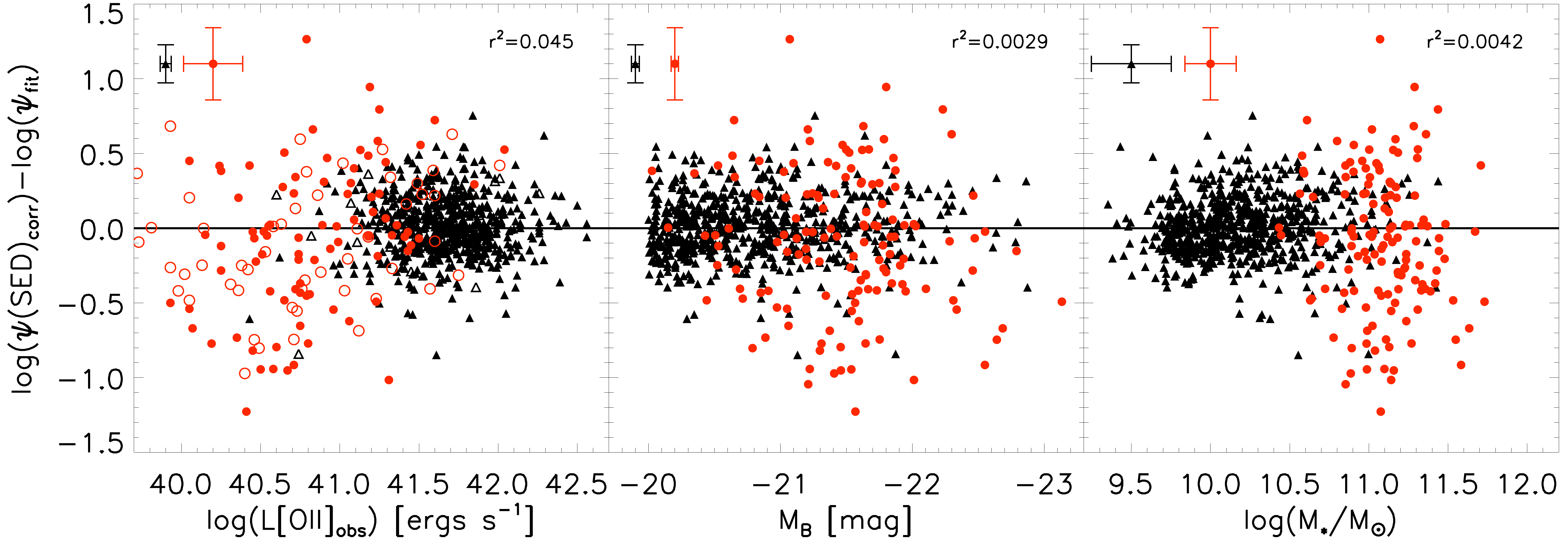}\\
\caption{SFR fit residuals using MB, (U-B), and (B-V) as the independent parameters in the SFR calibration for all galaxies. The residual errors are tabulated in Table~\ref{tab:SFRtotresults}. }
\label{fig:MBUBBV}
\end{figure*}
Knowing that additional color information can produce an accurate SFR calibration, we also generate a fit to the SED SFRs using M$_B$, (U-B), and (B-V) colors (see Figure~\ref{fig:MBUBBV}). The fit residuals in this ``multicolor" calibration are very similar to the second-order (U-B) calibration, with $\sigma_{all}=0.27$ dex scatter and small residual correlations in all three measurement parameters ($r^{2}<0.045$).  Again, we see that an additional color parameter allows the model more freedom to decouple the intrinsic galaxy color and the effects of dust in the SFR. The combined sample SFR calibration coefficients and fit residual errors for each of these fit models is tabulated in Table~\ref{tab:SFRtotresults}.

We note that it is not entirely surprising that a useful empirical SFR calibration can be achieved using observed $B$-band luminosity and multiple restframe colors; these measures are similar to the quantities that construct the SED-fit SFRs. Either the multicolor or second-order (U-B) calibration can be used to produce statistically similar SFR calibrations within our volume-limited ($0.74<z<1$) DEEP2 / AEGIS sample. However, we will see in the following section the multicolor calibration is better behaved in the $M_{B}$ - (U-B) plane when extended to the full $0.74<z<1.4$ DEEP2 sample.

\subsection{Trends in M$_B$ - (U-B) plane}
\label{sec:contours}
\begin{figure}[tb]
\centering
\includegraphics[width=0.98\columnwidth]{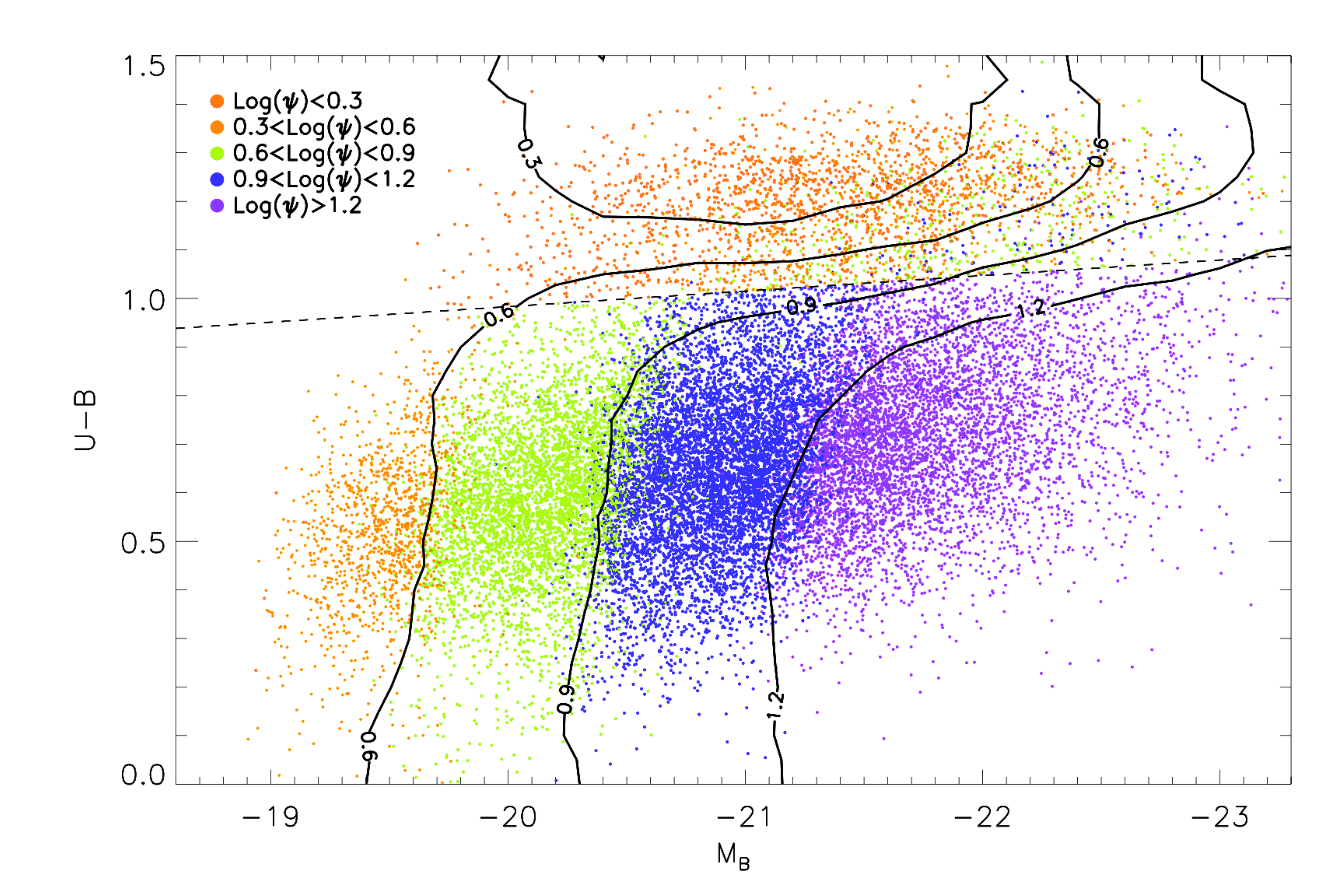}
\caption{The diagnostic M$_{B}$ - (U-B) plane with SFRs calibrated using the M$_{B}$ - L[\ion{O}{2}]  calibration split between red and blue galaxies (dotted line). The solid lines correspond to constant contours of SFR generated from the matched S09 AEGIS galaxy sample with log($\psi$)=[0.3, 0.6, 0.9, 1.2] M$_{\sun}$ yr$^{-1}$ and are plotted here for easy reference to the original calibration data. See Section~\ref{sec:RedBlueCal} for details of the color-split SFR calibration.}
\label{fig:2colorSFRcont}
\end{figure}

An additional comparison diagnostic of our SFR calibration models can be observed in the M$_{B}$ - (U-B) plane. Figure~\ref{fig:2colorSFRcont} shows the SFRs calibrated from M$_{B}$ and L[\ion{O}{2}] using the bimodal restframe color-selected calibration detailed in Section~\ref{sec:RedBlueCal}. The galaxy data points are taken for the full DEEP2 sample from $0.74<z<1.4$ and are color-coded with $\Delta$log($\psi$)=0.3 dex SFR bins. The overplotted solid lines correspond to contours of constant SFR with log($\psi$)=[0.3, 0.6, 0.9, 1.2] M$_{\sun}$ yr$^{-1}$ from the S09 AEGIS sample (c.f. Figure~\ref{fig:S09SFRcontours}). In general, the color-separated SFR calibration tracks the S09 contours well, particularly in the blue galaxies where the SFR dependence is strongly correlated with M$_B$ and independent of restframe color. However, the drastic split between the galaxy colors near the green valley is clearly not representative of the true SFR transition at (U-B)$\sim$1. Additionally, bright galaxies just below the (U-B)=1 line have overpredicted SFRs relative to the S09 contours while bright red galaxies just redward of the line have underpredicted SFRs.  Figure~\ref{fig:S09SFRcontours} shows that there is some ``mixing" of SFRs for the brightest galaxies with optically red restframe colors, indicating that galaxies with (U-B)$>1$ at these redshifts can either be old early-type galaxies with low SFR or heavily-reddened late-type galaxies with a young stellar population and high SFR.
 
\begin{figure*}[tb]
\centering
\subfigure{\includegraphics[width=0.98\columnwidth]{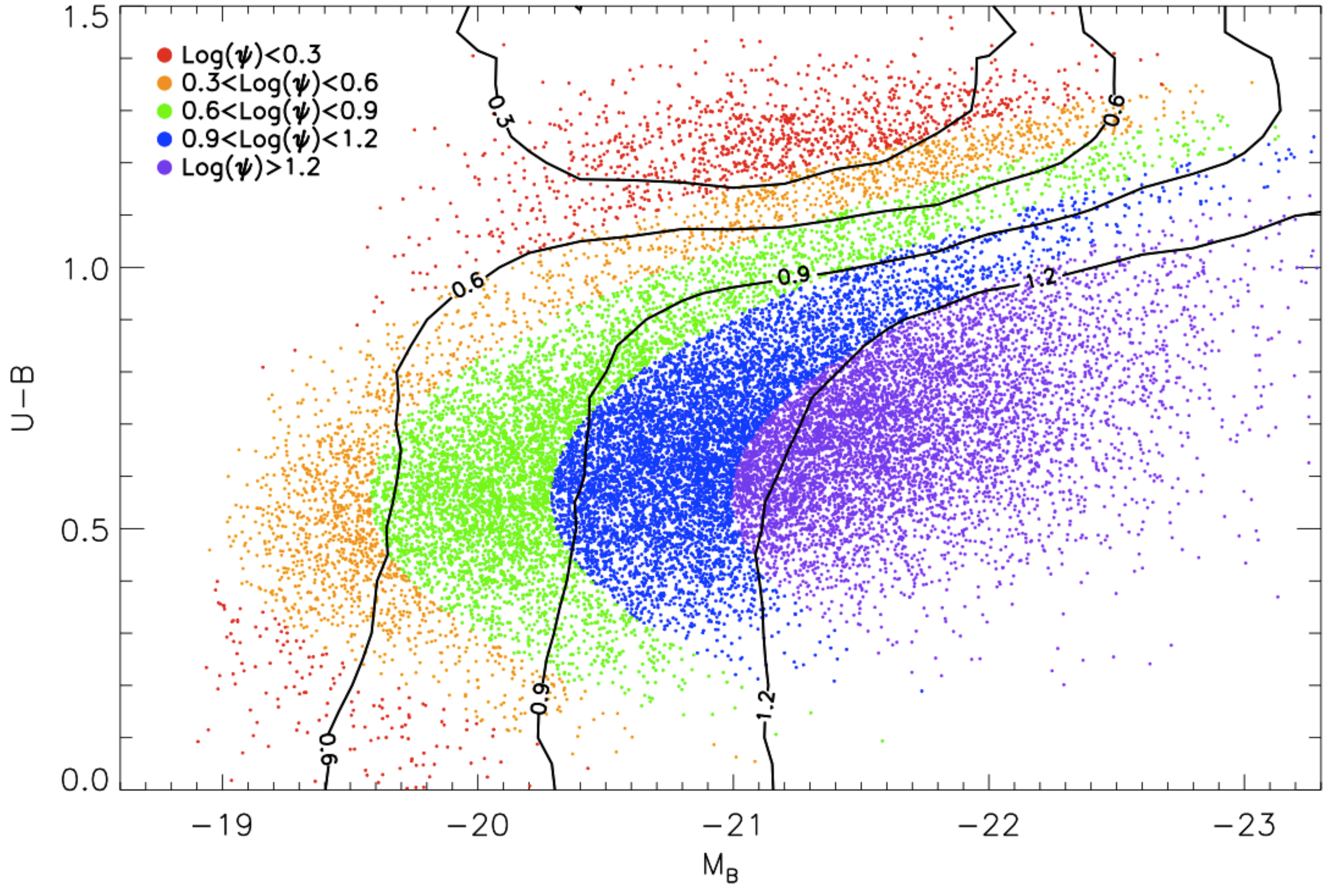}}
\subfigure{\includegraphics[width=0.98\columnwidth]{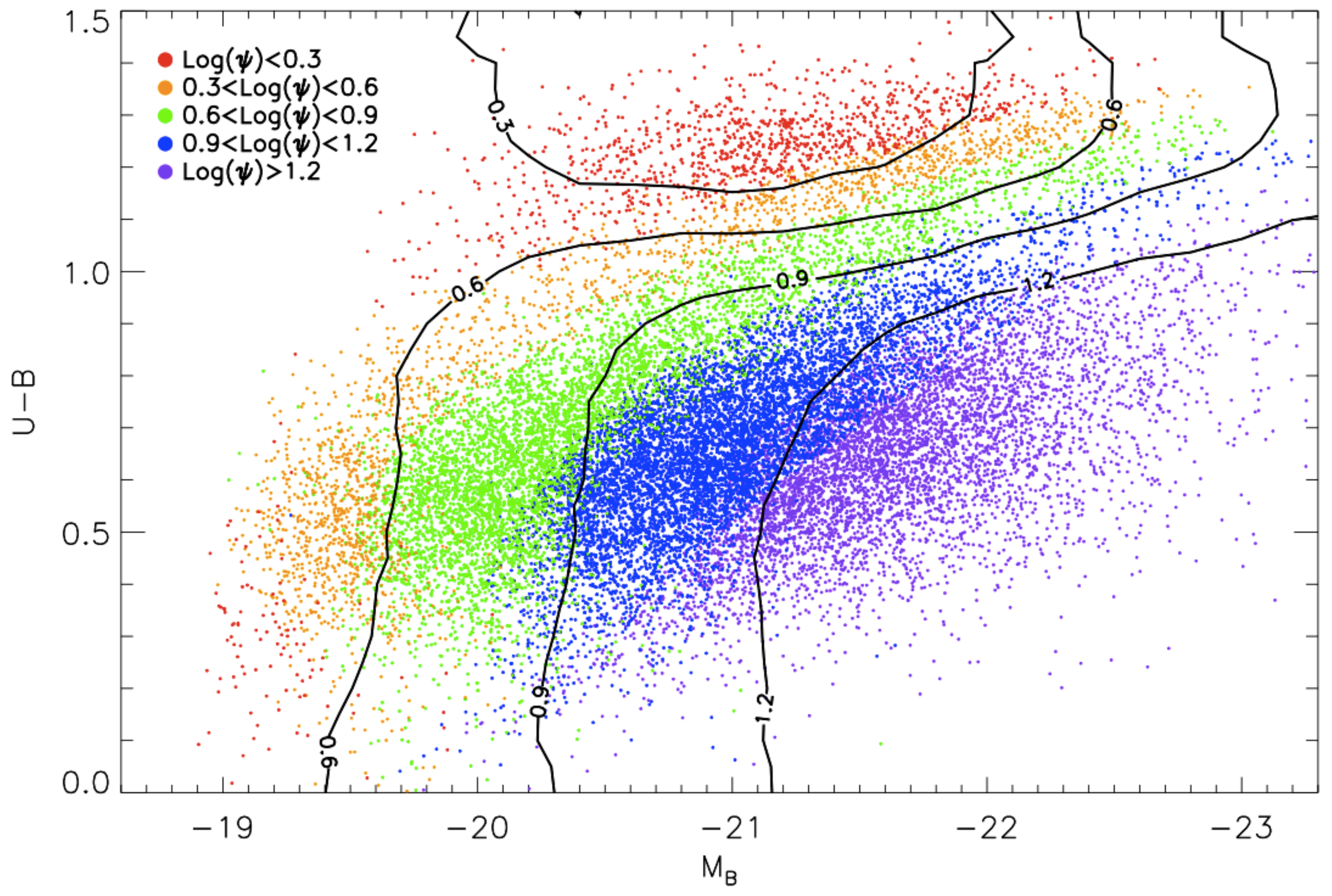}}
\caption{The diagnostic M$_{B}$ - (U-B) plane with calibrated $0.74<z<1.4$ DEEP2 SFRs using M$_{B}$ and a second-order polynomial in (U-B) restframe colors (left) and the SFR calibration using M$_{B}$, (U-B), and (B-V) restframe colors (right).}
\label{fig:SFRcontours}
\end{figure*}

As developed in Section~\ref{sec:MBUBcal}, the SFR transition from red to blue galaxies can be \emph{approximated} by a second-order polynomial fit in M$_{B}$ and (U-B) color that is valid within the range of SFRs and errors considered in our matched DEEP2 / AEGIS sample. The left hand plot of Figure~\ref{fig:SFRcontours} shows the calibrated SFRs for the full DEEP2 sample from $0.74<z<1.4$ using the [M$_{B}$, (U-B), (U-B)$^2$] calibration parameter model (see Table~\ref{tab:SFRtotresults}). Compared to the S09 SFR contours, the global SFR trends are reproduced by this calibration and SFRs smoothly transition between galaxies of differing restframe color. However, there is some discrepancy for faint blue galaxies; the SFRs become increasingly dependent on (U-B) color for (U-B)$<0.5$ and M$_{B}>-20$. This unphysical behavior is a consequence of the second-order term in the fit where there are few matched S09 data points to constrain the SFR calibration. 

The right hand plot of Figure~\ref{fig:SFRcontours} shows the linear SFR calibration using the multicolor calibration with M$_B$, (U-B), and (B-V) color used as independent fit parameters. Similar to the second-order M$_{B}$ - (U-B) SFR calibration, the multicolor calibration breaks some of the degeneracy in intrinsic galaxy color and dust reddening and therefore produces a smooth transition in (U-B) color between red and blue galaxies. However, the multicolor calibration remains insensitive to restframe colors for (U-B)$<1$ and therefore does not produce an unphysical turnover in the SFR dependence for faint blue galaxies.  We also note that multicolor SFR calibration produces increased SFRs for the brightest red galaxies (M$_{B}<-22.5$ and (U-B)$>1$). The S09 data shows that bright star-forming galaxies with heavily-reddened restframe colors do exist in this region. However, because the transition from high to low SFR is a strong function of (U-B) color, the SFRs in this region will not be highly accurate on an individual galaxy basis, but the SFR trend for the entire galaxy population will on average be correct. 

\begin{figure}[tb]
\centering
\includegraphics[width=0.98\columnwidth]{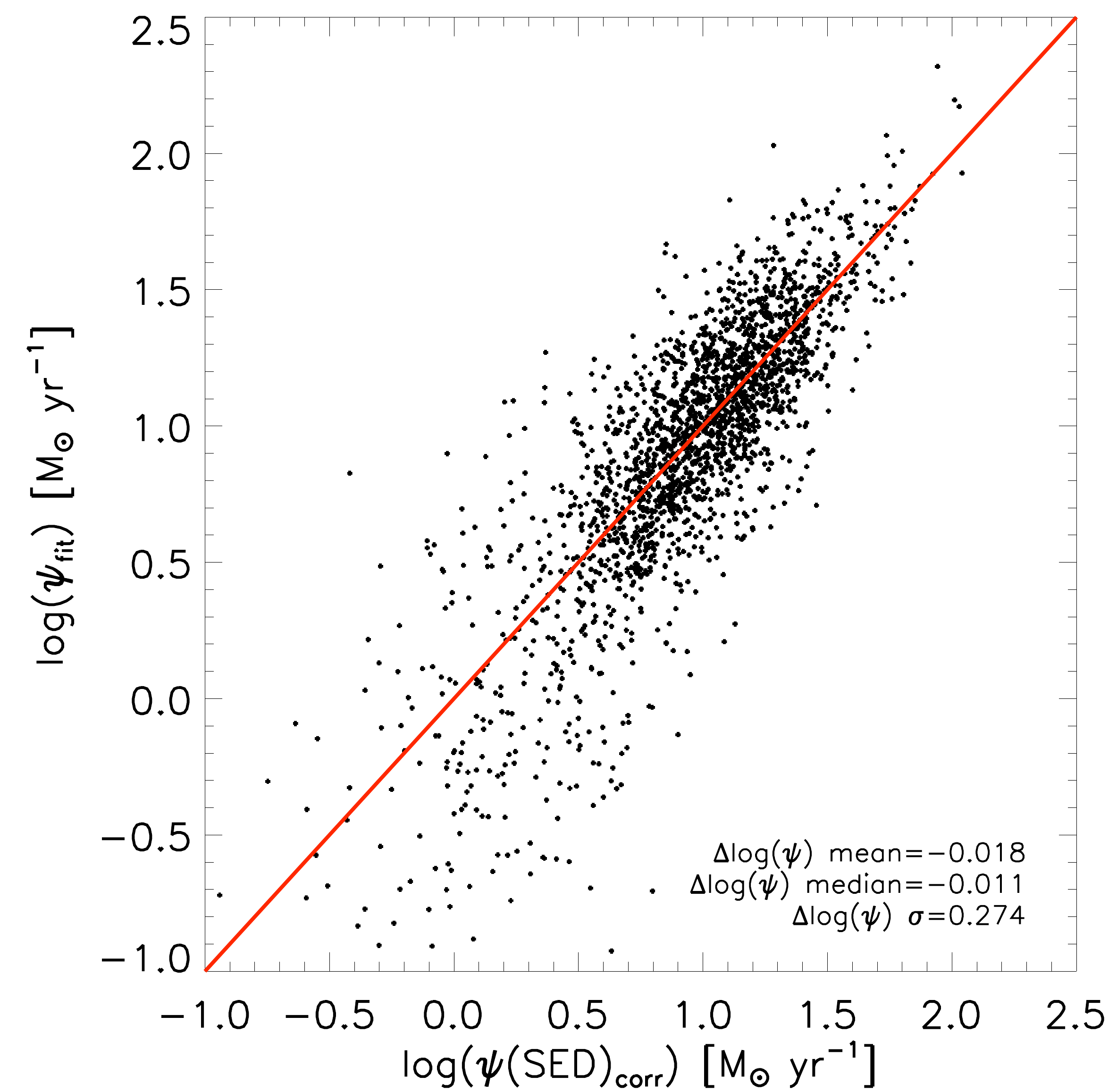}
\caption{SFR fit residuals using MB, (U-B), and (B-V) as the independent parameters in the SFR calibration for all DEEP2 galaxies from $0.74<z<1.4$.}
\label{fig:SFRfit}
\end{figure}

While the two calibrations used in Figure~\ref{fig:SFRcontours} are statistically equivalent in comparison to the S09 matched sample, we opt for the linear multicolor calibration to produce better qualitative agreement for faint blue galaxies. Figure~\ref{fig:SFRfit} shows the SFR produced by the multicolor calibration matches the ``true" SED SFRs very well, with a residual scatter of $\sigma_{all}=0.28$ dex for all galaxies with no offset in the mean of the distribution. In fact, the scatter is slightly reduced relative to the residual scatter seen in the original volume-limited calibration generated between $0.74<z<1$. The additional high redshift galaxies from $1<z<1.4$ are predominantly blue galaxies which have a more accurate SFR calibration than red galaxies through $M_{B}$.  The additional blue galaxies reduce the average residual scatter for the combined galaxy sample. 

\section{Comparison to local [\ion{O}{2}] SFR calibrations}
\label{sec:OIISFRs}
The [\ion{O}{2}] emission line has long been studied as a diagnostic for star formation in active galaxies \citep{Gallagher89, Hogg98, Kennicutt98, Teplitz03}. To remove the effects of dust extinction and oxygen abundance / metallicity, [\ion{O}{2}]  is typically calibrated to H$\alpha$ emission in the local universe, where H$\alpha$ luminosity is directly proportional to the number of ionizing photons available from massive stars.  The H$\alpha$ luminosity is also more robust against systematic error than other emission lines as it is generally insensitive to the ISM metallicity, experiences less dust attenuation, and absorption due to the underlying stellar continuum can be easily modeled \citep{MK06}. A calibration of the [\ion{O}{2}]/H$\alpha$ ratio can be readily performed to good accuracy at low redshifts where both lines can be simultaneously observed \citep[M06]{Kewley04}. At higher redshifts, the H$\alpha$ line typically becomes difficult to measure as it moves to NIR wavelengths, leaving estimates of the galaxy SFR dependent on assumptions of the mean [\ion{O}{2}]/H$\alpha$ ratio, extinction, and metallicity of lower-redshift samples. 

In this section, we study two empirical [\ion{O}{2}]-based SFR calibrations measured in the local universe and compare the results to the multicolor $z\sim1$ SFR calibration developed in \S\ref{sec:MBUBcal}. We reiterate that these empirical [\ion{O}{2}] SFR diagnostics are measured from large galaxy samples and are intended to represent the mean trends of the galaxy population. Applying the SFR calibrations on a individual galaxy-by-galaxy basis may lead to large variations and systematic error when compared to individual galaxy SFR diagnostics. Additional measures of dust extinction and metallicity should be obtained to accurately calculate an individual galaxy SFR.  

\subsection{[\ion{O}{2}] SFR calibration through M$_B$}
\label{sec:M06cal}
To mitigate the systematic effects that limit [\ion{O}{2}] as a SFR tracer, M06 uses an empirical calibration of the [\ion{O}{2}] SFR against the restframe $B$-band magnitude, which is correlated with both the galaxy extinction and metallicity. M$_B$ therefore serves as a proxy for the variation in the [\ion{O}{2}]/H$\alpha$ ratio from the local average value.  The M06 calibration produces good agreement (0.28 dex RMS scatter) with H$\alpha$-based SFRs at low redshift. M06 performed a comparison with galaxies at higher redshift ($z\sim0.7-1.5$) and found that the dust-corrected [\ion{O}{2}]/H$\alpha$ ratio remains roughly constant, although brighter galaxies tend to have lower ratios (see Fig. 22 of M06). The trend of lower observed [\ion{O}{2}]/H$\alpha$ ratios for brighter and more massive star-forming galaxies has been further confirmed with NIR measurements and tied to increased reddening through independent measurements of the extinction \citep{Weiner07}. To obtain SFRs for our sample, we follow the recommendation of M06 and interpolate between the median SFR values (their Table 2) using L[\ion{O}{2}] as a tracer of the SFR and M$_B$ to correct the SFR for dust extinction. The absolute $B$-band magnitudes assumed in M06 are in the Vega system, but we convert them to AB magnitudes for consistency within this study.

\begin{figure}[tb]
\centering
\includegraphics[width=0.98\columnwidth]{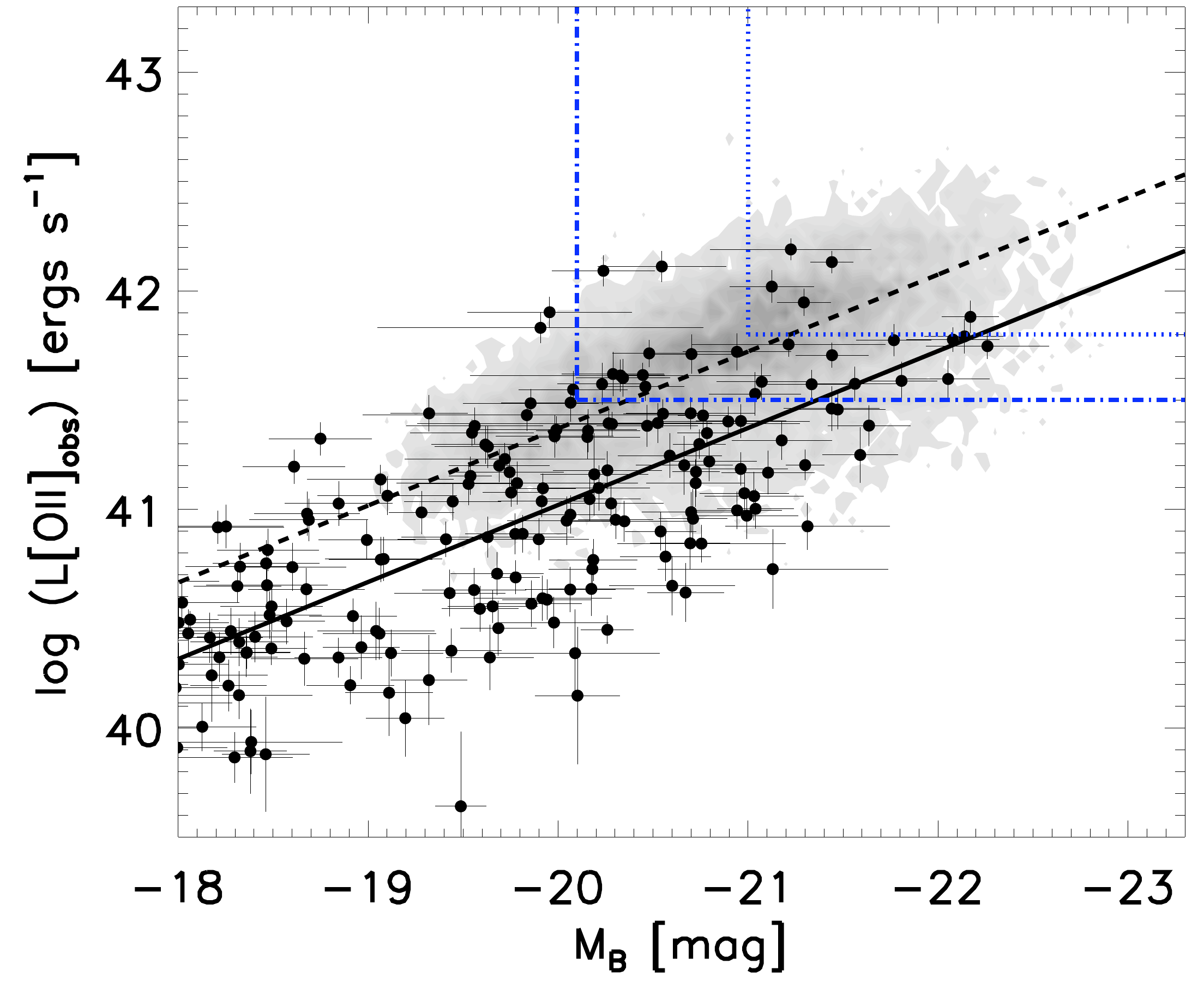}
\caption{The observed [\ion{O}{2}] emission line luminosity as a function of restframe $B$-band luminosity for the local sample measured in M06 (black triangles) and for DEEP2 galaxies from $0.74<z<1.4$ (gray contours). The vertical and horizontal lines mark the volume limits for blue DEEP2 galaxies at $z=1$ (dot-dashed) and $z=1.4$ (dotted). The solid (black) line is a weighted linear fit to the M06 data in the plotted range, and the dashed line is the same fit with a L[\ion{O}{2}] zeropoint shifted by +0.35 dex.}
\label{fig:MBvsOII}
\end{figure}
We stress that one should not $directly$ apply the locally-derived M06 SFR calibration using both observed L[\ion{O}{2}] and restframe M$_{B}$ measured for galaxies at intermediate redshift. Applying the local SFR relations and dust corrections to galaxies at intermediate redshifts can introduce considerable systematic uncertainty into the SFR calculation. It is well known that the $B$-band luminosity function has evolved since $z=1$ \citep{Willmer06, Faber07}, where the mean M$_B$ for integrated galaxy samples has dimmed by $Q\approx1.3$ magnitudes between $z=1$ and $z=0$. To show the discrepancy between the M06 local SFR calibration and that of a similar L[\ion{O}{2}] and M$_{B}$ relation constructed at $z\sim1$, we plot in Figure~\ref{fig:MBvsOII} the raw M06 data for star-forming local galaxies (M$_B<-18$) and $0.74<z<1.4$ blue galaxies from DEEP2. The blue lines in the figure correspond to the approximate volume limits of the DEEP2 measurements at $z<1$ (dot-dashed line) and $z<1.4$ (dotted line). For reference, we plot the weighted linear fit to the M06 data set with a black solid line, showing the trend of the galaxy continuum luminosity to be correlated with the observed [\ion{O}{2}] luminosity and therefore SFR for blue galaxies. Clearly, the DEEP2 data extends to higher luminosities in both L[\ion{O}{2}] and M$_{B}$ than M06, making the application of a SFR calibration at intermediate redshift heavily dependent on an extrapolation from the local values. It is also clear that the DEEP2 blue galaxy samples lie well above the mean local trend, however it is not immediately apparent whether the DEEP2 luminosities are brighter due to an evolution in M$_{B}$, L[\ion{O}{2}], or both.

Assuming \emph{no} evolution in M$_{B}$, the dashed line in Fig.~\ref{fig:MBvsOII} shows that the aggregate sample of DEEP2 blue galaxies has a $\sim0.35$ dex increase in observed [\ion{O}{2}] luminosity, and one would incur a similar bias in the mean SFR of $z\sim1$ galaxies. A more accurate calibration of the SFR at $z\sim1$ using the M06 data should at minimum account for the evolution in galaxy luminosity per fixed stellar mass. We therefore correct the measured values M$_B$ from DEEP2 galaxies to include a dimming factor of $Q=1.3$ magnitudes per unit redshift before applying the local M06 calibration. This simple luminosity correction factor implicitly assumes that both the [\ion{O}{2}]/H$\alpha$ ratio and stellar mass (or the mass-metallicity relation) has evolved less than the galaxy luminosity, an assumption in agreement with recent studies (\citealp{KK04, Zahid11}; J. Moustakas et. al, 2011, in preparation).  
\\
\subsection{[\ion{O}{2}] SFR calibration through Stellar Mass}
\label{sec:G10cal}
Recent efforts to use L[\ion{O}{2}] as a SFR diagnostic have focused on incorporating stellar mass into the calibration. Stellar mass has been shown to be correlated with many galaxy properties, including luminosity, metallicity, dust extinction, and the SFR in SDSS galaxies \citep{Brinchmann04, Blanton09, Garn10}. The stellar mass estimate serves as a proxy for these many complex factors and therefore accounts for the systematic differences in the observed [\ion{O}{2}]/H$\alpha$ ratio.  Following the formalism of \citet[hereafter G10]{Gilbank10}, we construct an empirical calibration of the [\ion{O}{2}] SFR from \citet{Kennicutt98} using
\begin{equation}
{\rm SFR_{L[O~\Rmnum{2}]} }= \frac{10^{0.4A_{H\alpha}}}{{[O~\Rmnum{2}]_{obs}}/[H{\alpha}]_{obs}}\frac{L[O~\Rmnum{2}]}{1.4\times10^{-41} {\rm erg~s}^{-1}}
\label{eq:Gilbank}
\end{equation}
in log(M$_{\sun}$ yr$^{-1})$, where {[O~\Rmnum{2}]$_{obs}$/[H${\alpha}$]$_{obs}$ is the average reddened line flux ratio between [\ion{O}{2}] and H$\alpha$ for the total sample of galaxies and $A_{H\alpha}$ is the extinction of H$\alpha$ line emission due to dust.  In the case of DEEP2 galaxies, we use a ratio of observed [\ion{O}{2}]/H$\alpha$=0.48 measured indirectly through [\ion{O}{2}]/H$\beta$ from $0.73<z<0.87$ and H$\beta$/H$\alpha$ from $0.33<z<0.39$ at M$_H$=-21 \citep{Weiner07}.  The value of  [\ion{O}{2}]/H$\alpha$ is in agreement with local measurements found in \citet{Kennicutt92}, but we caution that this value is not singular and can vary greatly depending on the galaxy sample selection. For the dust attenuation $A_{H\alpha}$, we use the mass-dependent dust extinction model of \citet{GB10}, who performed a PCA analysis of SDSS DR7 galaxies using SFR, stellar mass, and metallicity to produce a best fit model of the H$\alpha$ absorption. \citeauthor{GB10} found that a simple polynomial model in stellar mass best described the dust extinction using  
\begin{equation}
A_{H\alpha} = 0.91+ 0.77 X + 0.11X^2 - 0.09 X^3, 
\end{equation}
where $X$ = log(M$_{*}$/$10^{10}$ M$_{_{\sun}}$). Implicitly, this model accounts for deviations from the observed [\ion{O}{2}]/H$\alpha$ ratio and the observed differences in metallicity and luminosity that are correlated with the dust extinction. The data used to construct the model cover a large dynamic range of stellar mass for ``normal" galaxies, $8.5<{\rm log}(\rm M_{*}/\rm M_{sun})<11.5$, which spans the majority of the $z\sim1$ DEEP2 sample. 

\subsection{SFR Comparisons}
\label{sec:SFRcomp}

\subsubsection{SED-fit SFRs compared to L[\ion{O}{2}] SFRs}
To understand how the M06 and G10 calibrations compare to actual SFRs measured at $z\sim1$, we again use the S09 SFR data from our volume-limited DEEP2 / AEGIS sample. Because these calibrations were derived from local samples of star-forming galaxies, we restrict the sample to only galaxies with blue restframe colors and convert all SFRs to a Salpeter IMF for these comparisons.  

\label{sec:S09vsOII}
\begin{figure}[tb]
\centering
\includegraphics[width=0.98\columnwidth]{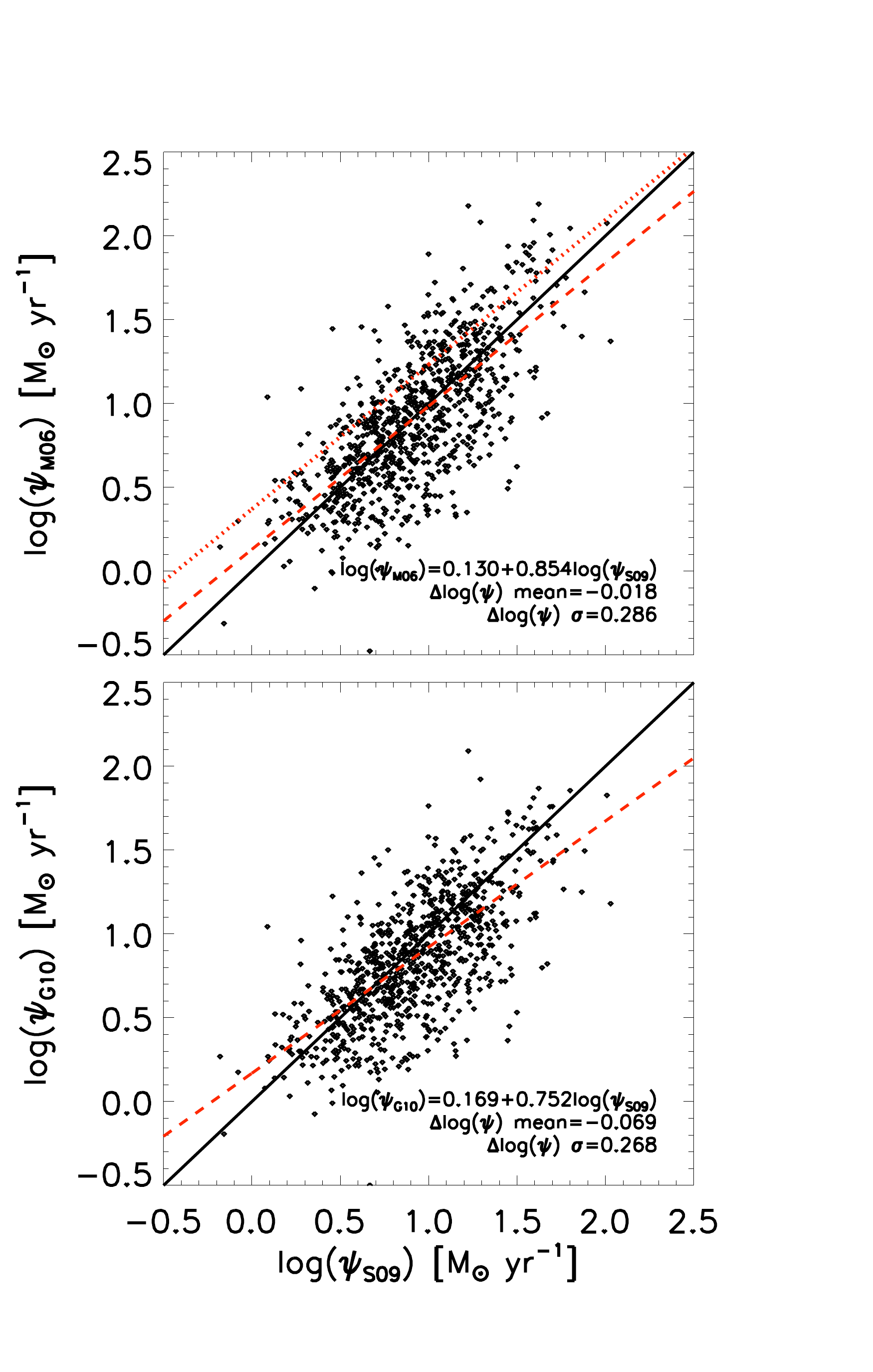}
\caption{Comparison of the S09 SFRs to calculated SFRs using L[\ion{O}{2}]  measurements and empirical dust corrections from the current epoch. The dashed (red) lines show the weighted linear least squares fit of the data points and the solid (black) line is the 1:1 correspondence. The top plot shows the S09 SFRs compared to the M06 empirical SFR relation using L[\ion{O}{2}] and M$_{B}$. The M06 calibration has been adjusted by $Q=1.3$ magnitudes per unit redshift in M$_B$ to account for the luminosity evolution at $z\sim1$. The dotted (red) line shows the M06 calibration before the luminosity evolution is taken into account.  The bottom plot shows the S09 SFRs and the G10 SFRs which uses L[\ion{O}{2}] and stellar mass for calibration. }
\label{fig:OIISFRdiff}
\end{figure}
Figure~\ref{fig:OIISFRdiff} shows the comparison between the L[\ion{O}{2}]-based SFR calibrations (M06 and G10) and the S09 SFRs from $0.74<z<1.0$. The dashed lines in the figure show the linear least squares fit of the sample weighted by the estimated errors in both measures of the SFR.  The upper panel of the plot shows the M06 calibration produces a 0.29 dex RMS residual scatter relative to the S09 data with a mean offset of -0.02 dex. The RMS scatter measured from the M06 calibration extrapolated to our sample is 0.08 dex larger than the residual error generated from the blue galaxy SFR calibration using L[\ion{O}{2}] and M$_{B}$ (see \S\ref{sec:RedBlueCal}). These results indicate that extrapolating the SFR behavior through L[\ion{O}{2}] and M$_{B}$ from the local relations to $z\sim1$ can contribute to the residual scatter, which is not unexpected given the systematic uncertainty in the galaxy dust corrections and average $B$-band luminosity evolution. We also plot a dotted line to represent the best fit trend of the M06 calibration without accounting for the luminosity evolution between $z=0.88$ and the current epoch. The difference between the SFR distribution means with and without luminosity evolution results in a bias of 0.25 dex. 

The bottom panel of Figure~\ref{fig:OIISFRdiff} shows the comparison of the G10 calibrated SFRs to the S09 SFRs.  We find that the SFR distributions agree with 0.27 dex RMS scatter and have a mean offset of -0.06 dex. While the mean offset is larger than the M06 calibration, the scatter is slightly smaller in the G10 case. Remarkably, the relation between stellar mass and $A_{H\alpha}$ in Equation~\ref{eq:Gilbank} produces a relatively accurate calibration at $z\sim1$ without having to account for any evolution. This result indicates that the dust extinction correction required for [\ion{O}{2}]-based SFRs is better correlated and more constant with stellar mass than restframe M$_{B}$. 

We note that the slopes of the weighted linear fits shown in Figure~\ref{fig:OIISFRdiff} are not well matched to a 1:1 correspondence. This is due in part to the large amount of scatter ($\sim0.28$ dex) and SFR measurement error relative to the overall SFR range probed ($\sim1.5$ dex) in the volume-limited blue galaxy sample. In order to better test the overall accuracy of the [\ion{O}{2}] SFR calibration slope, we must increase the sample size and probe a wider range of SFRs.

\subsubsection{Multicolor SFRs compared to L[\ion{O}{2}] SFRs}
\label{sec:MYSFRvsOII}
\begin{figure}[tb]
\centering
\includegraphics[width=0.98\columnwidth]{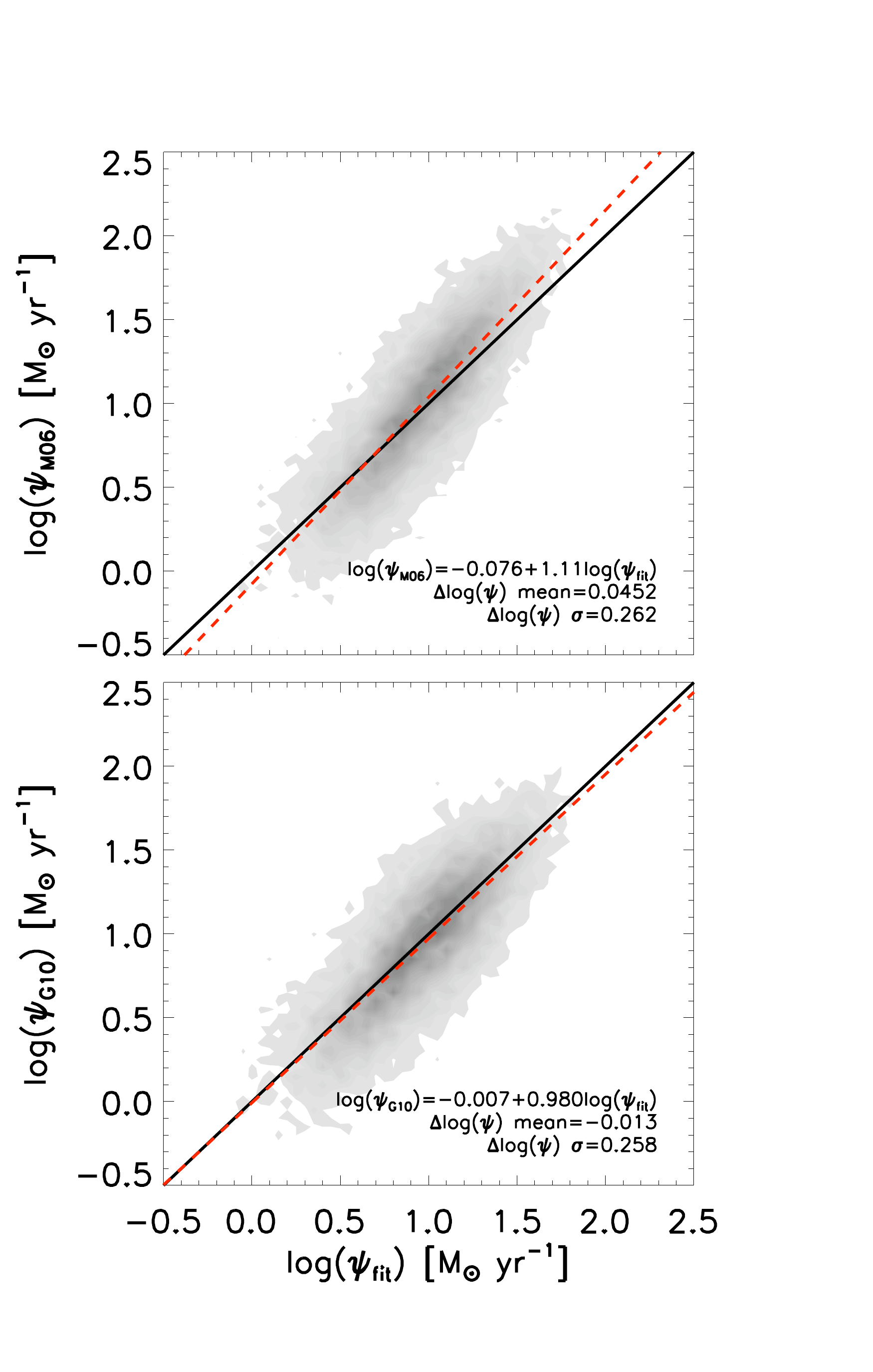}
\caption{Comparison of the M06 and G10 L[\ion{O}{2}]-based SFR calibrations to the multicolor SFR calibration for $0.74<z<1.4$ DEEP2 galaxies. The dashed (red) lines show the weighted linear regression of the data points and the solid (black) line is the 1:1 correspondence. Similar to Figure~\ref{fig:OIISFRdiff}, the M06 calibration has been adjusted by $Q=1.3$ magnitudes per unit redshift in M$_B$ to account for the luminosity evolution at $z\sim1$. }
\label{fig:SFRdiff}
\end{figure}
We now compare the multicolor SFR calibration presented in \S\ref{sec:MBUBcal} to the M06 and G10 SFR calibrations applied at intermediate redshifts, where we have extended the calibrations to include both red and blue galaxies from $0.74<z<1.4$ with measured L[\ion{O}{2}].  As mentioned in \S\ref{sec:Galtrends}, determining a SFR for a red galaxy based on the observed [\ion{O}{2}] luminosity can have large systematic uncertainty. However, there are some applications for which determining a rough SFR for large galaxy samples can still produce interesting global measures, e.g in measuring how the SFR correlates with environment \citep{Cooper06}. It is also interesting to check for systematic biases in [\ion{O}{2}] SFR calibrations when they are applied to red galaxies at $z\sim1$. 

As shown in the top panel of figure~\ref{fig:SFRdiff}, we find that the M06 empirical calibration from L[\ion{O}{2}] and M$_B$ is a reasonable match to our multicolor SED-fit SFR calibration \emph{after} a correction is applied for luminosity evolution in M$_B$ (see \S\ref{sec:M06cal}). The residual median offset between the SFR calibrations is 0.05 dex with a RMS residual error or 0.26 dex after the offset is removed.  The residual RMS scatter has been reduced slightly over the previous volume-limited blue galaxy sample because we have added in many star-forming blue galaxies at $z>1$. The SFR calibration slope differs by 11\% per decade, perhaps due to additional systematic bias such as an evolution in the [\ion{O}{2}]/H$\alpha$ ratio between $z=1$ and the current epoch. Alternatively, the disagreement could be due to differential luminosity evolution as our assumption of a constant $Q=1.3$ mag/$z$ corresponds primarily to $L_{*}$ galaxies.  

The bottom panel of Figure~\ref{fig:SFRdiff} shows that the G10 calibration, modified to correct dust extinction in L[\ion{O}{2}]  through the stellar mass, is an excellent match to our multicolor SFR calibration. The mean of the SFR distributions have a very small offset of 0.01 dex and the slope of the SFRs are in agreement to better than 2\% per decade.  Overall, both the M06 and G10 SFR calibrations seem to be generally consistent with our multicolor SFR calibration, an encouraging result considering the multicolor calibration is based only on broadband optical colors and restframe magnitude. We do not observe a significant systematic bias in the red galaxy SFRs estimated through either L[\ion{O}{2}]-based SFR calibrations relative to our multicolor SFR calibration.

Finally, we compare the SED-fit SFR contours in the restframe color-magnitude plane to those generated from the M06 and G10 calibrations. As shown in Figure~\ref{fig:OIISFRcontours}, both L[\ion{O}{2}]-based calibrations produce similar behavior as the S09 SFR contours in the color-magnitude plane, but the G10 calibration is slightly better at reproducing the color-independent SFR behavior seen in the blue cloud and therefore is in better agreement with our multicolor calibration for blue galaxies (c.f. Figure~\ref{fig:SFRdiff}). Combined with the previous comparison to the SED-fit SFRs, these results show that our SFR calibration using only restframe optical colors and $B$-band luminosity are at least as accurate as using a local L[\ion{O}{2}]-based SFR relation and extrapolating it to $z\sim1$.  

\begin{figure*}[thb]
\centering
\subfigure{\includegraphics[width=0.98\columnwidth]{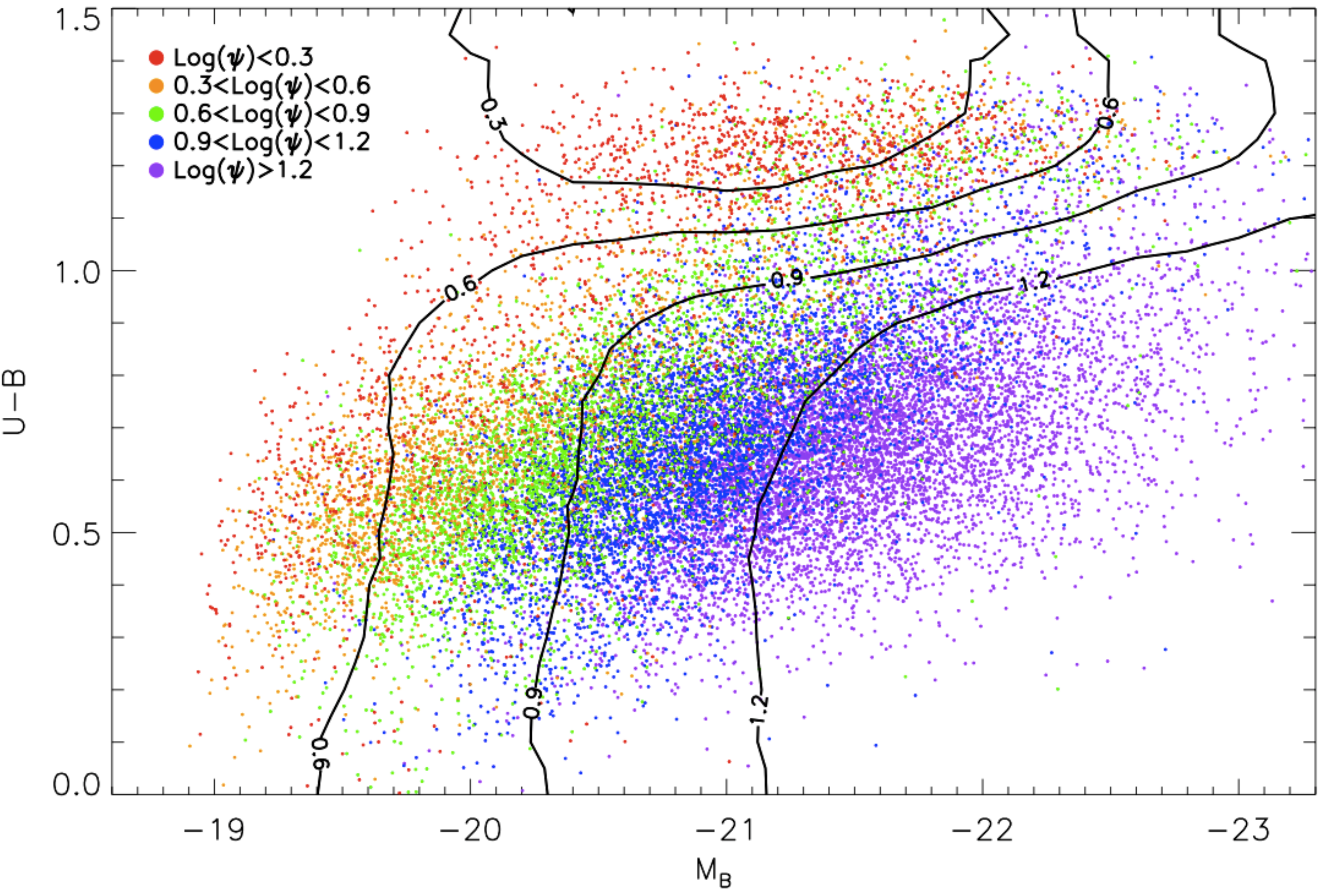}}
\subfigure{\includegraphics[width=0.98\columnwidth]{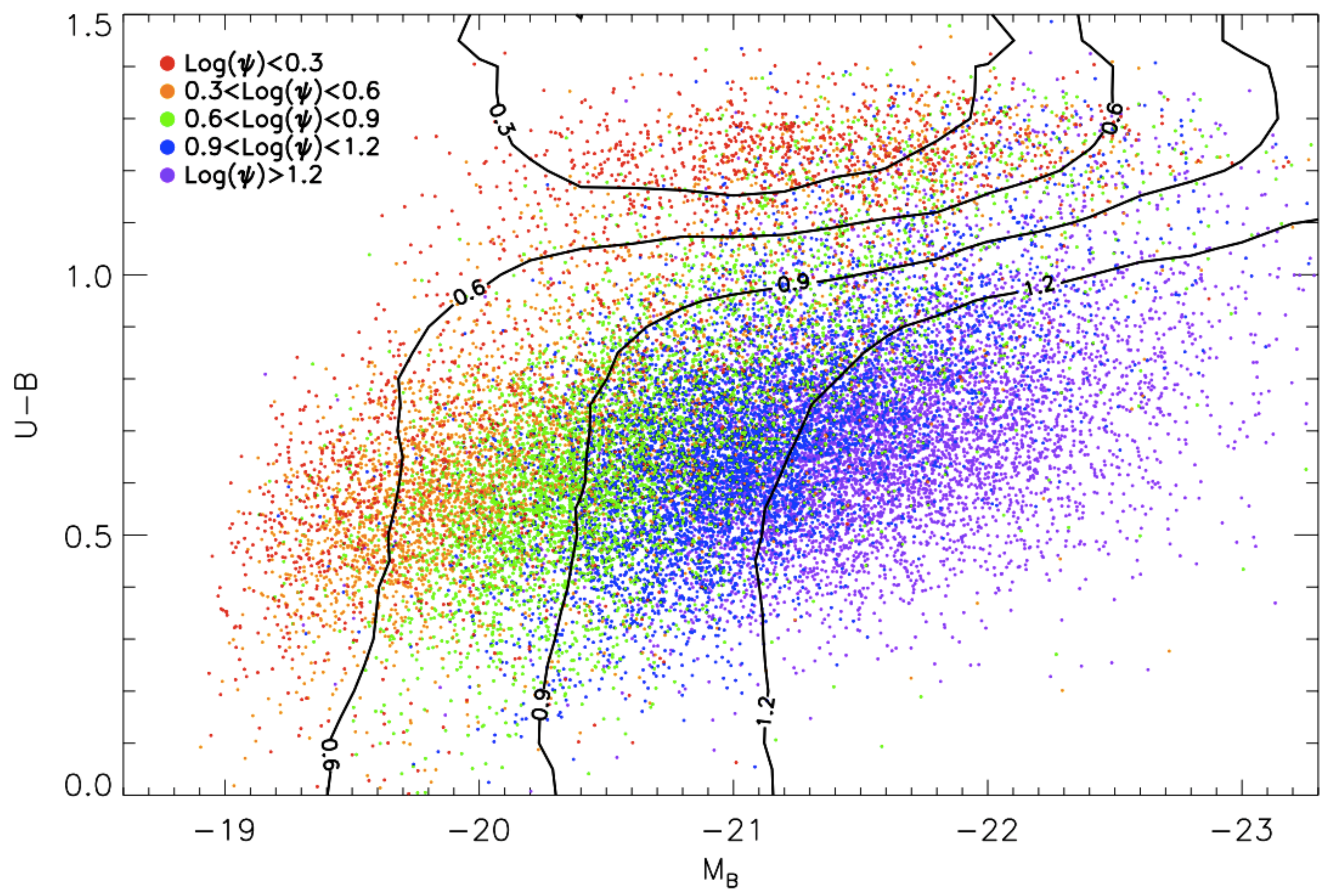}}
\caption{The M$_{B}$ - (U-B) plane with calibrated $0.74<z<1.4$ DEEP2 SFRs using the M06 empirical SFR calibration (left) and the G10 empirical SFR calibration (right) based on local SFR measurements.  The contour lines correspond to the S09 SFRs of log($\psi$)=[0.3, 0.6, 0.9, 1.2] M$_{\sun}$ yr$^{-1}$.}
\label{fig:OIISFRcontours}
\end{figure*}
 
\section{Conclusions}
\label{sec:Conclusions}
We present a calibration of the star formation rate for DEEP2 galaxies using SED-fit SFRs from AEGIS and observed correlations in [\ion{O}{2}], M$_{B}$, and restframe colors. Within a volume-limited sample of galaxies between $0.74<z<1.0$, we formulate and minimize a weighted $\chi^2$ fit with respect to these parameters  to determine a simple, robust empirical SFR calibration. This calibration is useful for estimating the global SFR behavior for a large sample of optically-measured galaxies at $z\sim1$, but it is not recommended for use on an individual galaxy basis. We summarize our main results as follows. 

\begin{enumerate}
\item{Red sequence and blue cloud galaxies at $z\sim1$ have large differences in their SFRs. As such, an accurate SFR calibration must minimally separate the galaxies into two populations based on restframe color {\it or} include the restframe color into the calibration. }
\item{We find that a single linear relation in M$_{B}$ and restframe (U-B) and (B-V) colors breaks the observed degeneracies in the intrinsic galaxy color and dust reddening and results in a SFR calibration within the accuracy of the available data. The calibration produces fit residual errors with 0.27 dex RMS scatter for the total sample ($\sigma_{blue}=0.20$ dex, $\sigma_{red}=0.47$ dex) and no significant systematic residual correlations with L[\ion{O}{2}], M$_{B}$, or the color-M/L stellar mass. Additional constraint from observed [\ion{O}{2}] luminosity does not improve the calibration accuracy, likely due to the fact that the SED-fit SFRs are highly correlated with the galaxy UV/optical continuum for blue galaxies and L[\ion{O}{2}] is less correlated with star formation in red galaxies.} 
\item{The multicolor SFR calibration developed from the matched DEEP2 / AEGIS sample can be extended to all DEEP2 galaxies from $0.74<z<1.4$ without a significant increase in SFR residual scatter. The fit SFR behavior in the M$_B$ - (U-B) color plane accurately reflects the average galaxy population in the S09 data without introducing significant discontinuities between red and blue galaxies.} 
\item{We compare the SED-fit SFRs to two calibrations (M06 and G10) which use local [\ion{O}{2}] luminosity measurements as a primary indicator of SFR and apply empirical dust corrections through observed M$_B$ or stellar mass. We find that the RMS scatter between the calculated SFRs are 0.29 dex (M06) and 0.27 dex (G10), which is $\sim0.08$ dex larger than the residual scatter generated from the best SFR calibrations in this study for the same volume-limited blue galaxy sample.}
\item{The empirical L[\ion{O}{2}] - M$_B$ SFR calibration found in M06 should be corrected for the luminosity evolution in star-forming galaxies when applied at $z\sim1$.  Applying the locally-measured M06 relation at intermediate redshift and not correcting for the mean galaxy luminosity evolution will overestimate the $z=1$ SFRs by 0.3 dex or more.  An L[\ion{O}{2}]-based SFR calibration that uses stellar mass to correct for the dust extinction (G10) does not require a large correction for evolution, indicating that stellar mass is better correlated with extinction than $M_{B}$ and has less evolution from $z\sim1$ to the current epoch.}
\item{Finally, we compare the multicolor SFR calibration recommended in this study to the L[\ion{O}{2}]-based SFR calibrations from M06 and G10. We find that our empirical calibration is in good agreement with both of the L[\ion{O}{2}] calibrations for all galaxies within the SFR calibration accuracy. We conclude that the multicolor SFR calibration fit from the AEGIS SED-fit SFRs is equivalent in accuracy to the local L[\ion{O}{2}] SFR calibrations extrapolated to $z\sim1$.}
\end{enumerate}

We gratefully thank the DEEP2 and AEGIS team for making their data public. Funding for the DEEP2 survey has been provided by NSF grants AST95-09298, AST-0071048, AST-0071198, AST-0507428, and AST-0507483 as well as NASA LTSA grant NNG04GC89G. Some of the data presented herein were obtained at the W. M. Keck Observatory, which is operated as a scientific partnership among the California Institute of Technology, the University of California and the National Aeronautics and Space Administration. The Observatory was made possible by the generous financial support of the W. M. Keck Foundation. The DEEP2 team and Keck Observatory acknowledge the very significant cultural role and reverence that the summit of Mauna Kea has always had within the indigenous Hawaiian community and appreciate the opportunity to conduct observations from this mountain. This study makes use of data from AEGIS, a multiwavelength sky survey conducted with the $Chandra$, $GALEX$, $Hubble$, Keck, CFHT, MMT, $Subaru$, Palomar, $Spitzer$, VLA, and other telescopes and supported in part by the NSF, NASA, and the STFC. This work was also sponsored in part by the United States Department of Energy under contract No. DE-AC02-05CH11231.

\clearpage
\begin{sidewaystable*}[p]
\caption{Summary of best fit coefficients, coefficient error, and fit residual errors for the full sample SFR calibration}
\begin{center}
\begin{tabularx}{0.98\textwidth}{llllccccccccccccccc}
\hline
\hline
\multicolumn{4}{c}{Fit Parameters\T} & \multicolumn{5}{c}{Coefficients} & \multicolumn{5}{c}{Coefficient Error} & Intrinsic Scatter$^a$ & \multicolumn{4}{c}{Fit Residuals$^a$ } \\
$P_{1}$\B & $P_{2}$ & $P_{3}$ & $P_{4}$~~&~~$C{_0}$ & $C{_1}$  & $C{_2}$ & $C_{3}$ & $C_{4}$~~&~~$\sigma_{0}$ & $\sigma_{1}$ & $\sigma_{2}$ & $\sigma_{3}$ & $\sigma_{4}$~~&~~$\sigma_{i}$~~&~~$\sigma_{red}$ & $\sigma_{blue}$ & $\sigma_{all}$ & $\Delta_{mean}$ \\
\hline
M$_{B}$\T & (U-B) & -- & -- & 2.039 & -0.408 & -1.436 & -- & --  & 0.043 & 0.020 & 0.054 & -- & -- & 0.27 & 0.48 & 0.25 & 0.33 & -0.01\\[3pt]
M$_B$ & (U-B) & (U-B)$^2$ & -- & 0.381 & -0.424  & 2.925 & -2.603 & -- & 0.082 &  0.015 & 0.206 & 0.122 & -- & 0.19 &  0.47 & 0.19 & 0.26 & -0.02\\[3pt]
M$_B$ \B& (U-B) & (B-V) & -- & 1.199 & -0.431 & -11.251 & 15.446  & -- & 0.055 & 0.016 & 0.526 & 0.821 & -- & 0.20 & 0.47 & 0.20 & 0.27 & -0.01\\[3pt]
\hline
L[\ion{O}{2}]\T & (U-B) & -- & -- & 0.536 & 0.705 & 0.005  & -- & -- & 0.064 & 0.035 & 0.062 &  -- & -- & 0.24 & 0.45 & 0.28 & 0.33 & -0.02\\[3pt]
L[\ion{O}{2}] & (U-B) & (U-B)$^2$ & -- & -0.290 & 0.599 & 2.50 & -1.601 & --  & 0.097 & 0.034 & 0.245 & 0.154 & -- & 0.24 &  0.46 & 0.28 & 0.33 & -0.01\\[3pt]
L[\ion{O}{2}] \B& (U-B) & (B-V) & -- & 0.187 & 0.629 & -5.818 & 8.954 & -- & 0.068 & 0.034 & 0.641 & 0.977 & -- & 0.22 & 0.44 & 0.27 & 0.31 & -0.01\\[3pt]
\hline
L[\ion{O}{2}]\T & M$_B$ & (U-B) & -- & 1.203  & 0.420 & -0.284 & -0.637  & -- & 0.073 & 0.037 & 0.0202 & 0.073  & -- & 0.20 & 0.44 & 0.23 & 0.30 & -0.02\\[3pt]
L[\ion{O}{2}] & M$_B$ & (U-B) & (U-B)$^2$ & 0.270 & 0.157 & -0.381  & 2.721 & -2.325 & 0.079 & 0.034 & 0.017 & 0.193 & 0.128 & 0.14  & 0.44 & 0.20 & 0.26 & 0.00\\[3pt]
L[\ion{O}{2}] \B& M$_B$ & (U-B) & (B-V) & 0.921 & 0.217 & -0.367 & -9.309 & 12.990 & 0.066 & 0.0.035 & 0.018 & 0.562 & 0.833 & 0.16 & 0.43 & 0.20 & 0.26 & -0.01\\[3pt]
\hline
\hline
\end{tabularx}
\end{center}
\hspace{0.8in}$^a$ SFR error is in units of dex.
\label{tab:SFRtotresults}
\end{sidewaystable*}
\clearpage


\begin{thebibliography}{}

\bibitem[Bell et al.(2003)]{Bell03} Bell, E.~F., McIntosh, 
D.~H., Katz, N., \& Weinberg, M.~D.\ 2003, \apjs, 149, 289 

\bibitem[Blanton \& Moustakas(2009)]{Blanton09} Blanton, 
M.~R., \& Moustakas, J.\ 2009, \araa, 47, 159 

\bibitem[Brinchmann et al.(2004)]{Brinchmann04} Brinchmann, J., 
Charlot, S., White, S.~D.~M., Tremonti, C., Kauffmann, G., Heckman, T., 
\& Brinkmann, J.\ 2004, \mnras, 351, 1151 

\bibitem[Bruzual 
\& Charlot(2003)]{BC03} Bruzual, G., \& Charlot, S.\ 2003, \mnras, 344, 1000 

\bibitem[Bundy et al.(2006)]{Bundy06} Bundy, K., et al.\ 2006, 
\apj, 651, 120 

\bibitem[Calzetti et al.(1994)]{Calzetti94} Calzetti, D., Kinney, 
A.~L., \& Storchi-Bergmann, T.\ 1994, \apj, 429, 582

\bibitem[Calzetti(2001)]{Calzetti01} Calzetti, D.\ 2001, \pasp, 
113, 1449 

\bibitem[Calzetti et al.(2007)]{Calzetti07} Calzetti, D., et al.\ 
2007, \apj, 666, 870 

\bibitem[Calzetti(2009)]{Calzetti09} Calzetti, D.\ 2009, Cosmic 
Dust - Near and Far , 414, 214

\bibitem[Charlot 
\& Fall(2000)]{CF00} Charlot, S., \& Fall, S.~M.\ 2000, \apj, 539, 718 

\bibitem[Cooper et al.(2006)]{Cooper06} Cooper, M.~C., et al.\ 
2006, \mnras, 370, 198 

\bibitem[Coil et al.(2008)]{Coil08} Coil, A.~L., et al.\ 2008, 
\apj, 672, 153 

\bibitem[Daddi et al.(2007)]{Daddi07} Daddi, E., et al.\ 2007, 
\apj, 670, 156 

\bibitem[Dale 
\& Helou(2002)]{DH02} Dale, D.~A., \& Helou, G.\ 2002, \apj, 576, 159

\bibitem[Davis et~al.(2003)]{Davis03}
Davis, M., et~al.\ 2003, Proc.~of SPIE, 4834, 161.

\bibitem[Davis et al.(2007)]{Davis07} Davis, M., et al.\ 2007, 
\apjl, 660, L1 

\bibitem[Elbaz et al.(2007)]{Elbaz07} Elbaz, D., et al.\ 2007, \aap, 468, 33 

\bibitem[Faber et al.(2003)]{Faber03} Faber, S.~M., et al.\
2003, Proc.~of SPIE, 4841, 1657.

\bibitem[Faber et al.(2007)]{Faber07} Faber, S.~M., et al.\ 
2007, \apj, 665, 265 

\bibitem[Gallagher et al.(1989)]{Gallagher89} Gallagher, J.~S., 
Hunter, D.~A., \& Bushouse, H.\ 1989, \aj, 97, 700 

\bibitem[Garn \& Best(2010)]{GB10} Garn, T., \& Best, P.~N.\ 2010, \mnras, 409, 421 

\bibitem[Garn et al.(2010)]{Garn10} Garn, T., et al.\ 2010, 
\mnras, 402, 2017 

\bibitem[Gerke et al.(2007)]{Gerke07} Gerke, B.~F., et al.\ 
2007, \mnras, 376, 1425 

\bibitem[Gilbank et al.(2010)]{Gilbank10} Gilbank, D.~G., Baldry, 
I.~K., Balogh, M.~L., Glazebrook, K., 
\& Bower, R.~G.\ 2010, \mnras, 405, 2594 

\bibitem[Han et al.(2007)]{Han07} Han, Z., Podsiadlowski, P., 
\& Lynas-Gray, A.~E.\ 2007, \mnras, 380, 1098 

\bibitem[Heckman et al.(1998)]{Heckman98} Heckman, T.~M., Robert, 
C., Leitherer, C., Garnett, D.~R., 
\& van der Rydt, F.\ 1998, \apj, 503, 646 

\bibitem[Hogg et al.(1998)]{Hogg98} Hogg, D.~W., Cohen, J.~G., 
Blandford, R., \& Pahre, M.~A.\ 1998, \apj, 504, 622 

\bibitem[Hopkins 
\& Beacom(2006)]{Hopkins06} Hopkins, A.~M., \& Beacom, J.~F.\ 2006, \apj, 651, 142 

\bibitem[Kennicutt(1992)]{Kennicutt92} Kennicutt, R.~C., Jr.\ 1992, 
\apj, 388, 310 


\bibitem[Kennicutt(1998)]{Kennicutt98}
Kennicutt, Jr., R.~C.\ 1998, \araa, 36, 189.

\bibitem[Kennicutt et al.(2009)]{Kennicutt09} Kennicutt, R.~C., et 
al.\ 2009, \apj, 703, 1672 

\bibitem[Kewley \& Dopita(2002)]{KD02} 
Kewley, L.~J., \& Dopita, M.~A.\ 2002, \apjs, 142, 35 

\bibitem[Kewley, Geller \& Jansen(2004)]{Kewley02}
Kewley, L.~J., Geller, M.~J.\ \& Jansen, R.~A.\ 2004, \aj, 127, 2002.

\bibitem[Kewley et al.(2004)]{Kewley04} Kewley, L.~J., Geller, 
M.~J., \& Jansen, R.~A.\ 2004, \aj, 127, 2002 

\bibitem[Kobulnicky \& Kewley(2004)]{KK04} Kobulnicky, H.~A., \& Kewley, L.~J.\ 2004, \apj, 617, 240 

\bibitem[Kong(2004)]{Kong04} Kong, X.\ 2004, \aap, 425, 417

\bibitem[Moustakas, Kennicut \& Tremonti(2006)]{Moustakas06}
Moustakas, J., Kennicutt, Jr., R.~C.\ \& Tremonti, C.~A.\ 2006, \apj, 642, 775.

\bibitem[Moustakas \& Kennicutt(2006)]{MK06} 
Moustakas, J., \& Kennicutt, R.~C., Jr.\ 2006, \apjs, 164, 81 

\bibitem[Noeske et al.(2007a)]{Noeske07a} Noeske, K.~G., et al.\ 
2007, \apjl, 660, L43 %SFH paper

\bibitem[Noeske et al.(2007b)]{Noeske07b} Noeske, K.~G., et al.\ 
2007, \apjl, 660, L47  %stellar mass paper

\bibitem[Rieke et al.(2009)]{Rieke09} Rieke, G.~H., 
Alonso-Herrero, A., Weiner, B.~J., P{\'e}rez-Gonz{\'a}lez, P.~G., Blaylock, 
M., Donley, J.~L., \& Marcillac, D.\ 2009, \apj, 692, 556 

\bibitem[Salim et al.(2007)]{Salim07} Salim, S., et al.\ 2007, 
\apjs, 173, 267 

\bibitem[Salim et al.(2009)]{Salim09} Salim, S., et al.\ 2009, 
\apj, 700, 161 

\bibitem[Salpeter(1955)]{Salpeter55} Salpeter, E.~E.\ 1955, \apj, 
121, 161 

\bibitem[Steidel et al.(1999)]{Steidel99} Steidel, C.~C., 
Adelberger, K.~L., Giavalisco, M., Dickinson, M., 
\& Pettini, M.\ 1999, \apj, 519, 1 

\bibitem[Storey 
\& Hummer(1995)]{SH95} Storey, P.~J., \& Hummer, D.~G.\ 1995, \mnras, 272, 41 

\bibitem[Teplitz et al.(2003)]{Teplitz03} Teplitz, H.~I., 
Collins, N.~R., Gardner, J.~P., Hill, R.~S., 
\& Rhodes, J.\ 2003, \apj, 589, 704

\bibitem[Wang 
\& Heckman(1996)]{WH96} Wang, B., \& Heckman, T.~M.\ 1996, \apj, 457, 645 

\bibitem[Weiner et al.(2006)]{Weiner06} Weiner, B.~J., et al.\ 
2006, \apj, 653, 1049 %TF paper

\bibitem[Weiner et al.(2007)]{Weiner07} Weiner, B.~J., et al.\ 
2007, \apjl, 660, L39 

\bibitem[Weiner et al.(2009)]{Weiner09} Weiner, B.~J., et al.\ 
2009, \apj, 692, 187 

\bibitem[Willmer et al.(2006)]{Willmer06} Willmer, C.~N.~A., et 
al.\ 2006, \apj, 647, 853 

\bibitem[Yan et al.(2006)]{Yan06} Yan, R., Newman, J.~A., 
Faber, S.~M., Konidaris, N., Koo, D., \& Davis, M.\ 2006, \apj, 648, 281 

\bibitem[Zahid et al.(2011)]{Zahid11} Zahid, H.~J., Kewley, 
L.~J., \& Bresolin, F.\ 2011, \apj, 730, 137 

\bibitem[Zhu et al.(2008)]{Zhu08} Zhu, Y.-N., Wu, H., Cao, 
C., \& Li, H.-N.\ 2008, \apj, 686, 155 

\bibitem[Zhu, Moustakas \& Blanton(2009)]{Zhu09}
Zhu, G., Moustakas, J.\ \& Blanton, M.~R.\ 2009, \apj, 701, 86.

\end{thebibliography}
\end{document}